\RequirePackage{fix-cm}
\documentclass[smallextended]{svjour3}       
\smartqed  

\usepackage{graphicx}

\usepackage{acronym}

\usepackage{listings}
\usepackage[dvipsnames]{xcolor}
\usepackage{graphicx}
\usepackage[table]{xcolor}
\usepackage{multirow}
\usepackage[most]{tcolorbox}
\usepackage{tabularx}
\usepackage{array}
\usepackage{makecell}
\usepackage{amssymb}   
\usepackage{pifont}
\usepackage{tikz}
\usepackage{ragged2e}
\usepackage{acronym}
\usepackage[caption=false,font=footnotesize]{subfig}
\usepackage{pdflscape}
\usepackage{longtable}
\usepackage{booktabs}

\usepackage[strings]{underscore}

\definecolor{codegreen}{rgb}{0,0.6,0}
\definecolor{codegray}{rgb}{0.5,0.5,0.5}
\definecolor{codepurple}{rgb}{0.58,0,0.82}
\definecolor{backcolour}{rgb}{0.95,0.95,0.92}

\lstdefinestyle{mystyle}{
    backgroundcolor=\color{backcolour},
    commentstyle=\color{codegreen},
    keywordstyle=\color{blue},
    numberstyle=\tiny\color{codegray},
    stringstyle=\color{codepurple},
    basicstyle=\ttfamily\footnotesize,
    breakatwhitespace=false,         
    breaklines=true,                 
    captionpos=b,                    
    keepspaces=true,                 
    showspaces=false,                
    showstringspaces=false,
    showtabs=false,                  
    tabsize=2
}

\newtcolorbox{rqsummarybox}[1][]{
    colback=gray!10,
    colframe=gray!60,
    boxrule=0.5pt,
    arc=2pt,
    left=6pt,
    right=6pt,
    top=6pt,
    bottom=6pt,
    title=#1,
    fonttitle=\bfseries,
    coltitle=black,
    colbacktitle=gray!20,
    boxed title style={
        boxrule=0pt,
        sharp corners,
    },
    separator sign={\rule{\linewidth}{0.4pt}} 
}

\newcommand{\greencirclecheck}{%
\tikz[baseline=-0.6ex]{
    \fill[green!70!black] (0,0) circle (0.13cm);
    \node at (0,0) {\textcolor{white}{\scriptsize\checkmark}};
}}

\begin{document}
\title{Understanding Efficient LLM Serving: An Empirical Study of Frameworks, Optimizations Algorithms, and Real-World Adoption
}

\titlerunning{Understanding Efficient LLM Serving}        

\author{Forough Majidi         \and
        Mohammad Mehdi Morovati \and
        Foutse Khomh \and
        Heng Li
}


\institute{F. Majidi \and
            MM. Morovati \and
            F. Khomh \and
            H. Li
            \at
              SWAT Lab., Polytechnique Montréal, Montréal, Canada \\
\email{forough.majidi@polymtl.ca, mehdi.morovati@polymtl.ca, \\ foutse.khomh@polymtl.ca, heng.li@polymtl.ca}
}

\date{Received: date / Accepted: date}

\maketitle

\begin{abstract}
Large Language Models (LLMs) are integrated into software systems and AI services, making efficient LLM serving a concern for software engineering. Serving LLMs is challenging because inference requires computation, memory, GPU resources, and execution while maintaining latency and throughput. Although prior research has proposed LLM inference, optimization, and serving techniques and frameworks, little is known about how they are adopted in practice. In this study, we investigate the use of LLM serving frameworks and serving methods in open-source software systems. We identify and analyze five LLM-specific frameworks: vLLM, SGLang, TensorRT-LLM, LMDeploy, and FlashInfer. We examine how these frameworks and techniques are adopted individually and in combination, how adoption varies across categories of LLMs, and how repositories differ in intent, focus, use case, and architectural design. Our results show that vLLM is the most visible framework in popularity and adoption, while parallel computation, memory management, and network pruning are the most frequently used serving-method categories. Multi-framework usage is limited, suggesting that developers rely on a single serving framework; however, combined frameworks connect complementary capabilities across the serving stack. Framework adoption varies across model families, modalities, model sizes, domain specializations, and deployment settings, indicating differences in support for operational requirements. Repository-level analysis shows that LLM serving frameworks support applications and architectures, including Reinforcement Learning (RL)-based reasoning, multimodal generation and understanding, microservices, and cloud infrastructure. Overall, this study provides a large-scale empirical characterization of LLM serving framework adoption in practice and offers insights for researchers, framework maintainers, and practitioners working on LLM systems.

\end{abstract}

\section{Introduction}

Large Language Models (LLMs) are increasingly being integrated into software systems and user-facing AI services, driving rapid growth in applications such as conversational agents, code assistants, content generation platforms, and decision-support systems~\cite{pan2025survey,chang2024survey}. As these models transition from research prototypes to production systems, the efficiency of model deployment and serving has become a critical software-engineering concern. Serving LLMs requires managing substantial computational and memory demands while meeting strict latency, throughput, scalability, and cost requirements~\cite{miao2025towards}. Recent surveys highlight that efficient LLM serving depends not only on model execution but also on system-level mechanisms such as request scheduling, memory management, kernel optimization, and resource orchestration \cite{pan2025survey}. Consequently, LLM serving has emerged as a key systems challenge that directly influences the feasibility, performance, and cost-effectiveness of real-world AI applications. These challenges are further amplified by the diversity of deployment environments. Modern LLM services must operate across heterogeneous hardware platforms, cloud infrastructures, and workload characteristics, often under stringent resource and cost constraints \cite{jiang2025thunderserve}. During inference, memory consumption can become a major bottleneck, limiting throughput and increasing operational costs when not managed efficiently \cite{kwon2023efficient}. As a result, scalable LLM deployment increasingly relies on specialized serving frameworks that coordinate model execution, request processing, memory allocation, scheduling, and hardware-specific optimizations. To address these challenges, researchers and practitioners have proposed a growing ecosystem of LLM serving frameworks and optimization techniques. Frameworks such as vLLM, SGLang, TensorRT-LLM, LMDeploy, and FlashInfer encapsulate a wide range of capabilities, including memory-efficient attention mechanisms, scheduling strategies, kernel optimizations, model compression techniques, and hardware-aware execution engines~\cite{kwon2023efficient,zheng2024sglang,TensorRT_LLM_document,LMDeploy_document,FlashInfer_document}. These frameworks are rapidly becoming foundational infrastructure for deploying production-grade LLM applications.

Despite the rapid evolution of LLM serving technologies, our understanding of their adoption in practice remains limited. Prior work has extensively surveyed efficient inference techniques, model compression methods, deployment optimizations, and serving architectures~\cite{1bai2024beyond,9wan2023efficient,13xu2024survey,14zhou2024survey,pan2025survey,miao2025towards}. Other studies have proposed and evaluated specific algorithmic and systems-level optimizations, including quantization, speculative decoding, GPU-serving architectures, and production-serving infrastructures~\cite{12xia2024unlocking,10wang2024model,15zhu2024survey,3ding2024hybrid,4gao2024cost,jiang2025thunderserve,11xia2023flash,2beck2025evaluation,8shan2025aibrix}. However, little empirical evidence exists on how these frameworks and serving methods are actually adopted, combined, and deployed in real software systems. In particular, it remains unclear which serving frameworks have gained traction among developers, how efficient serving methods are used in practice, how frameworks and optimization techniques are combined, which frameworks are preferred for different categories of LLMs, and what types of software systems rely on these technologies.
Addressing this gap is important for both research and practice. Understanding real-world adoption patterns can reveal which innovations successfully transition from research into deployment, identify underexplored opportunities for framework development, and help practitioners make informed decisions about serving architectures for production systems. It can also provide framework maintainers with empirical insights into ecosystem needs, integration patterns, and emerging deployment requirements.

To address this gap, we conduct an empirical study of LLM serving framework adoption in open-source GitHub repositories. We investigate the adoption of serving frameworks, the efficient serving methods associated with them, framework and method co-usage patterns, framework adoption across different categories of LLMs, and the characteristics of software systems that rely on these frameworks. By connecting serving technologies with repository-level system characteristics, our study provides a comprehensive view of how efficient LLM serving is realized in practice and offers actionable insights for researchers, framework maintainers, and practitioners building production-grade LLM systems.

To guide this empirical study, we formulate three research questions. The first question examines which LLM serving frameworks and efficient serving methods are most popular and most adopted in practice. The second question investigates how efficient serving methods and serving frameworks are used in combination, as well as which serving frameworks developers adopt for different types of LLM. The third question examines the types of LLM-based software systems that adopt serving frameworks and how this adoption differs across repository intent, technical focus, primary use case, and system design.

\begin{description}
    \item[RQ1.] What are the most popular and widely adopted LLM serving frameworks and efficient serving methods?
    \begin{itemize}
        \item What are the most popular LLM serving frameworks?
        \item What are the most adopted LLM serving frameworks?
        \item What are the most used efficient LLM serving methods?
    \end{itemize}

    \item[RQ2.] What are the most used combinations of efficient LLM serving methods and serving frameworks, and which serving frameworks are adopted for different types of LLM?
    \begin{itemize}
        \item What are the most used combinations of efficient LLM serving methods?
        \item What are the most used combinations of LLM serving frameworks?
        \item What serving frameworks are adopted by developers for each type of LLMs?
    \end{itemize}

    \item[RQ3.] How are LLM serving frameworks used across different types of LLM-based software systems?
    \begin{itemize}
        \item What types of LLM-based software systems, in terms of repository intent, technical focus, use case, and system design, adopt LLM serving frameworks?
        \item Which types of LLM-based software systems most commonly adopt each LLM serving frameworks, and how does this adoption vary across repository intent, technical focus, primary use case, and system design?
    \end{itemize}
\end{description}

The results show that LLM serving is used unevenly across frameworks, methods, model contexts, and repository types. vLLM is the most visible framework in both popularity and adoption, while parallel computation, memory management, and network pruning are the most frequently observed serving-method categories, with framework-specific patterns such as the stronger role of kernel fusion in FlashInfer repositories. The results also show that developers sometimes combine multiple serving methods within the same framework, whereas multi-framework usage remains limited. In multi-framework repositories, vLLM has the largest absolute co-usage count, while FlashInfer has the highest co-usage ratio, suggesting that it is more often adopted alongside other serving frameworks. These combinations appear to connect complementary capabilities across the serving stack, including scheduling, memory management, parallel execution, and attention- or kernel-level optimization. For model usage, frameworks are associated with different model families, sizes, specializations, and deployment contexts, with vLLM and LMDeploy used with broader model-family ranges than some other frameworks. Finally, repository-level analysis shows that LLM serving frameworks are adopted across diverse contexts, including Mathematical Reasoning with RL, Speech, OCR, and Multimedia Processing Pipelines, Multimodal Generation and Understanding, and Microservices and Cloud Infrastructure.

This study makes the following contributions:
\begin{itemize}
    \item We provide an empirical characterization of LLM serving framework popularity and adoption in open-source GitHub repositories, focusing on LLM-specific serving frameworks selected from literature and GitHub sources.

    \item We identify and analyze the efficient serving methods used across the studied frameworks, showing how different categories of serving methods,  such as parallel computation, memory management, network pruning, request scheduling, quantization, and kernel optimization, are used in open-source repositories.

    \item We examine how LLM serving is used in combination by analyzing efficient serving method combinations within frameworks, serving-framework combinations across repositories, and the serving frameworks adopted for different types of LLM, including model families, modalities, sizes, specializations, and deployment contexts.

    \item We characterize the types of LLM-based software systems that adopt serving frameworks by clustering repository summaries and analyzing repository intent, technical focus, primary use case, and system design.

    \item We provide a replication package containing the materials and datasets used in the study to support transparency and reproducibility \cite{replicationPackage}.
\end{itemize}

The remainder of this paper is organized as follows. Section \ref{background} provides background on LLM and efficient LLM serving. Section \ref{sec:method} describes the methodology used to identify serving frameworks and methods, extract GitHub repositories, and characterize repository-level system designs. Section \ref{results} presents the empirical results for the research questions, including framework and method adoption, framework and method combinations, framework usage across LLM types, and repository-level system designs. Section \ref{discussion} presents the implications of our findings for researchers, framework maintainers, and developers. Section \ref{relatedworks} discusses related work on efficient LLM inference, serving methods, and deployment infrastructure. Section \ref{threats} discusses threats to validity. Finally, Section \ref{conclusion} concludes the paper and outlines future work.

\section{Background}
\label{background}
\subsection{Large Language Models }

AI broadly refers to computational methods that allow machines to carry out tasks commonly associated with human capabilities, including learning, reasoning, language understanding, and decision making \cite{ibm_ai_document,russell2021artificial,gignac2024defining}.
Within AI, generative AI has recently become a prominent area, focusing on the creation of new content such as text, code, images, audio, and video from user prompts \cite{feuerriegel2024generative,banh2023generative,ibm_generative_ai_document}.
Large Language Models (LLMs) represent one of the main classes of generative AI models. These models are generally trained on large collections of text and code and commonly rely on transformer-based architectures to understand, process, and generate natural language and other types of content \cite{naveed2025comprehensive,banh2023generative,ibm_llm_document}. As a result, LLMs have been applied to a wide range of natural language processing tasks, including text generation, summarization, machine translation, question answering, and code generation \cite{qin2026large,chang2024survey,naveed2025comprehensive,minaee2024large}. More recent multimodal extensions further connect language models with visual inputs, supporting applications that combine natural language processing and computer vision \cite{yin2024survey,li2025survey}. Beyond general-purpose applications, LLMs are also being adopted across a growing range of domain-specific settings. In sectors such as healthcare, finance, autonomous driving, insurance, and actuarial services, they are being explored to support complex tasks including medical question answering, clinical documentation and summarization, financial analysis, decision support, claims processing, and workflow automation~\cite{wang2024large,nie2024survey,zhu2024survey,balona2024actuarygpt,kong2024financial}. 

Although LLMs have achieved strong results in many language- and software-related tasks, using them in real applications is still challenging. In comparison with many earlier task-specific ML and DL systems, LLMs generally involve larger model sizes, higher memory demands, and greater computation during inference, making their serving costly and resource-intensive \cite{zhou2024survey,miao2025towards}. These issues are more critical in production environments, where systems need to handle many users while maintaining low latency and high throughput. For instance, LLM serving often relies on expensive hardware accelerators such as GPUs, and limited hardware capacity can make it difficult for service providers to satisfy performance requirements \cite{jiang2025thunderserve}. Memory management is another important bottleneck, since the key-value cache used during generation can increase dynamically for each request and limit how many requests can be processed together \cite{kwon2023efficient}. Therefore, efficient LLM serving depends on methods and frameworks that can reduce resource usage, improve inference efficiency, and support scalable deployment in practical software systems \cite{pan2025survey,miao2025towards}.
\subsection{Efficient Serving}
Efficient LLM serving refers to deploying large language models in a way that reduces inference latency, lowers computational and memory overhead, and enables scalable operation in real-world applications. This is particularly important because LLM deployment is often resource-intensive, especially in settings that require serving many concurrent users while maintaining low latency and high throughput~\cite{miao2025towards,pan2025survey,majidi2026efficient}. In practice, efficient serving improves response times for end users, enables more effective utilization of hardware resources, and reduces the operational cost of LLM-based systems~\cite{kwon2023efficient,miao2025towards,majidi2026efficient}. 

A range of techniques have been proposed to improve the efficiency of LLM serving. One prominent direction focuses on memory optimization, where system designs more carefully manage activation and KV-cache memory during generation, allowing more concurrent requests to be handled on the same hardware infrastructure~\cite{kwon2023efficient}. Another direction focuses on improving the execution of inference requests, such as optimizing request scheduling, batching strategies, and the underlying organization of model computations to better utilize available hardware resources~\cite{pan2025survey,miao2025towards}. These approaches are particularly valuable because they enhance the practical efficiency of LLM systems without altering the functionality or intended behavior of the applications themselves.

Beyond these system-level improvements, efficient serving has important broader implications. For end users, it enables faster and more reliable responses in applications such as chatbots, search engines, code assistants, and other LLM-powered services. For organizations, it reduces the infrastructure burden associated with large-scale deployment, particularly when relying on expensive accelerators such as GPUs~\cite{miao2025towards,kwon2023efficient}. From an environmental standpoint, inference efficiency is increasingly recognized as a key concern, since LLM serving contributes to significant energy consumption and associated carbon emissions. Recent studies have therefore begun to explore strategies for improving the carbon efficiency of LLM inference workloads~\cite{li2024sprout,argerich2024measuring}. Overall, efficient LLM serving is not only a systems optimization problem, but also a practical requirement for deploying LLM-based software in a scalable, cost-effective, and environmentally responsible manner.

\section{Methodology}
\label{sec:method}
\begin{figure}
    \centering
    \includegraphics[width=0.9\linewidth]{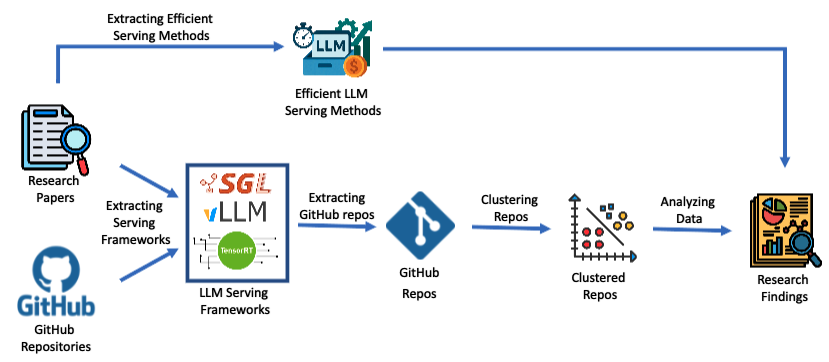}
    \caption{High level view of the methodology used for this study}
    \label{fig:method}
\end{figure}

This section presents the methodology adopted in this study. Fig.~\ref{fig:method} provides a high-level view of the overall research process. All materials and datasets used in this study are publicly available in the replication package~\cite{replicationPackage}.

\subsection{Extracting LLM Serving Frameworks}
\label{subsec:extract_serv_tool}
This study relies on two primary sources to identify LLM serving frameworks: scientific literature and GitHub repositories. To collect relevant research papers, we adopt a systematic search process commonly used in prior studies~\cite{majidi2022empirical}. 
Specifically, we query Google Scholar\footnote{https://scholar.google.ca} and Engineering Village\footnote{https://www.engineeringvillage.com} databases using a set of keywords related to cost efficiency and serving LLMs. It is worth noting that we limit our search to papers published after January 1, 2023, since cost efficiency and serving of LLMs only start to receive attention after that time. The keyword that we use in this step is \textit{("large language model" OR "LLM" OR "foundation model" OR "foundation") AND ("model serving" OR "model deployment" OR "inference serving" OR "serving" OR "deployment" OR "inference" OR "production") AND ("cost optimization" OR "cost reduction" OR "cost efficient" OR "cost effective" OR "cost aware" OR "resource optimization" OR "resource reduction" OR "efficiency improvement")}.
Next, we examine the top 100 results paper from each database, ranked by relevance. We read the whole text of retrieved papers and extract LLM serving or inference frameworks mentioned in the studied papers. Through this process, we identify 46 distinct frameworks reported in the literature for supporting LLM serving and efficiency experiments. The complete list of the identified frameworks from the research papers exist in the replication package of this paper \cite{replicationPackage}. 

To identify additional LLM serving frameworks that may not capture through our literature review, we conduct a complementary search on GitHub. Specifically, we use combinations of the keywords \texttt{`LLM'}, \texttt{`large language model'}, \texttt{`serving'}, and \texttt{`cost efficiency'} to retrieve relevant repositories. 
This process is conducted using the GitHub REST API v3~\cite{github_api_v3}. As a result, we initially identify 439 repositories.
Following prior studies~\cite{taraghi2026real,shah2026characterizing}, we filter out repositories with fewer than 500 stars to ensure dataset quality and exclude unpopular projects (based on the literature~\cite{li2025rise}),  
resulting in 26 repositories. Then, we manually examine the remaining repositories and exclude those that meet at least one of the following exclusion criteria.

\begin{itemize}
    \item Repositories with the descriptions in languages other than English.
    \item Repositories include only tutorials, examples, and training materials.
    \item Repositories without any implementation regarding efficiency and serving LLMs. 
\end{itemize}

\subsection{Extracting LLM Serving Methods}
\label{subsec:serv_method}
\begin{figure}
    \centering
    \includegraphics[width=0.9\linewidth]{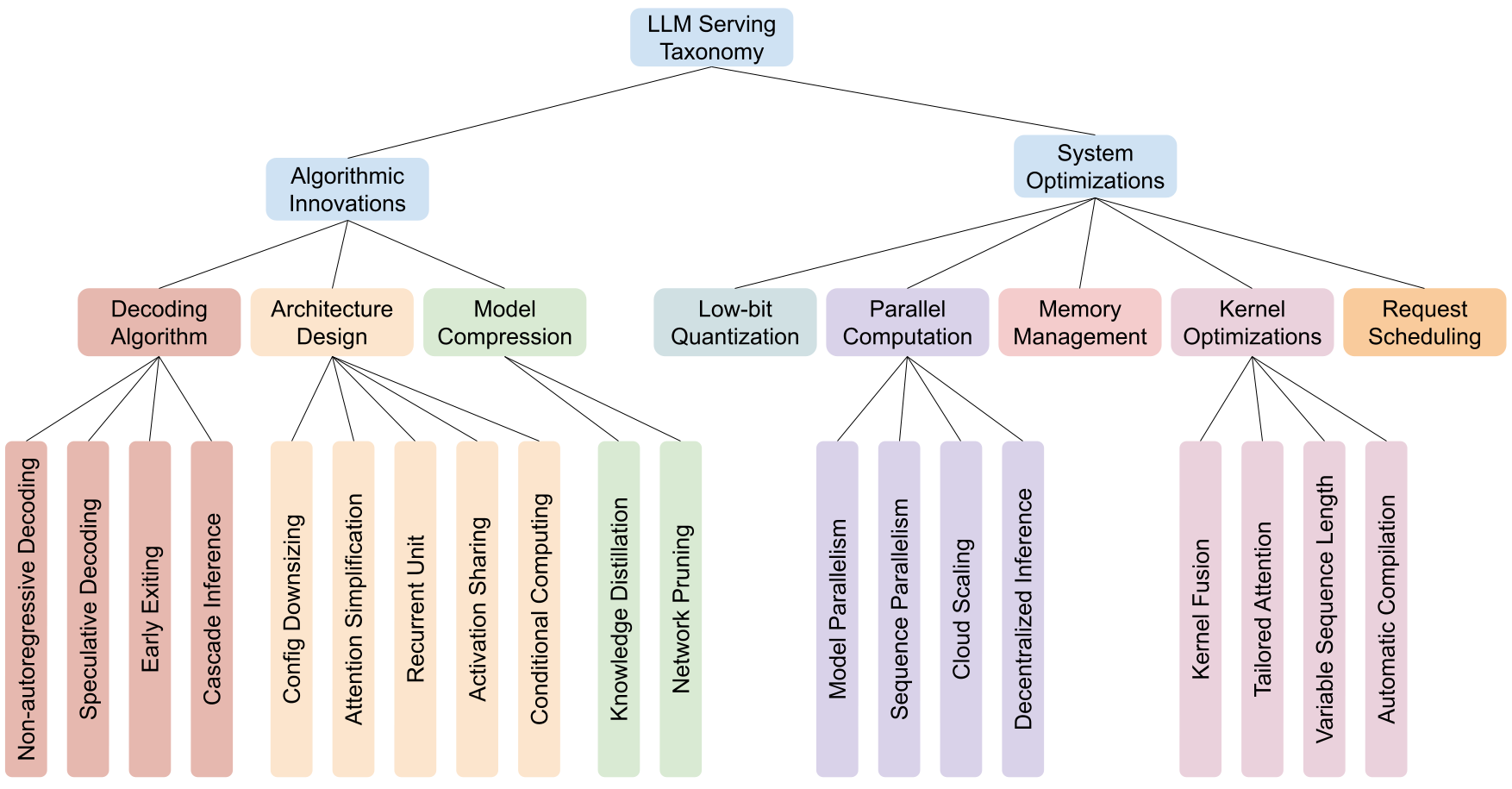}
    \caption{Taxonomy of existing methods on improving the LLM serving efficiency \cite{miao2025towards}.}
    \label{fig:servingmethods}
\end{figure}

To extract LLM serving methods, we manually review the full text of the research papers collected in the previous step (Subsection~\ref{subsec:extract_serv_tool}) and identify the techniques proposed to improve inference efficiency and serving performance. During this review, we record the serving methods described in each paper and map them to a common classification framework. Specifically, we adopt the taxonomy proposed by Miao et al.~\cite{miao2025towards}, which provides a comprehensive overview of existing approaches for enhancing LLM serving efficiency. As illustrated in Fig.~\ref{fig:servingmethods}, this taxonomy organizes serving techniques into two broad categories: (1) algorithmic innovations and (2) system-level optimizations~\cite{miao2025towards}. We use this taxonomy as a coding scheme to systematically classify the methods extracted from the literature and to facilitate a consistent comparison of approaches across studies.

In the subsequent step, we examine the documentation and technical materials of the serving frameworks identified earlier to determine which of the extracted serving methods are implemented by each framework. This allows us to analyze the adoption of different efficiency-enhancing techniques in existing LLM serving tools.


\subsection{Extracting Repositories}
\label{sec:method_extract_repos}
In order to analyze the adoption of LLM serving frameworks and serving methods identified in the previous steps (Section \ref{subsec:extract_serv_tool} and \ref{subsec:serv_method}), we collect open-source software systems that use these frameworks in practice. Following prior studies on software ecosystems and ML-based systems~\cite{shah2026characterizing,morovati2024bug}, we use GitHub as the primary source for identifying open-source projects. Since Python is the dominant programming language for developing LLM-based applications~\cite{morovati2024fault,taraghi2026real}, our analysis focuses exclusively on Python repositories.

To identify repositories that use the selected LLM serving frameworks, we adopt a methodology similar to those used in previous large-scale repository mining studies~\cite{morovati2023bugs,morovati2024fault}. First, for each framework, we review its official documentation and identify the Python APIs and import statements required to access its functionality. For example, developers typically use the statement \texttt{import vllm} to interact with the vLLM framework~\cite{vllm_document}. We then use the GitHub Search API~\cite{github_api_v3} to retrieve Python source files containing these framework-specific API usages.

Because the GitHub Search API returns at most 1,000 results per query, a single query may not capture all relevant files for popular frameworks. 
To overcome this limitation, we partition the search space using the file-size criterion (i.e., \texttt{size:<min>..<max>}), generating multiple non-overlapping queries that each return fewer than 1,000 results. 
As no relevant files exceeding 500 KB are identified during our preliminary investigation, we restrict our search to files within the 1 byte and 500 KB size range. Next, we divide this range into 500 bytes intervals, yielding approximately 1,000 queries per framework. This systematic querying strategy ensures the comprehensive collection of Python scripts that employ the targeted serving frameworks.
After collecting the candidate files, we identify the unique repositories associated with them and apply a set of exclusion criteria, following previous repository-mining studies~\cite{morovati2024bug,morovati2023bugs}, to remove repositories that are unlikely to represent mature and actively maintained software projects. Specifically, we exclude:

\begin{itemize}
\item \textbf{Unpopular repositories:} repositories with fewer than 10 stars and fewer than 10 forks.
\item \textbf{Inactive repositories:} repositories with no recorded activity during the previous year.
\item \textbf{Personal repositories:} repositories with only one contributor.
\item \textbf{Repositories with trivial history:} repositories containing fewer than 100 commits.
\end{itemize}

To study the adoption of specific serving methods, we further analyze the Python scripts extracted from the selected repositories. Because a single LLM serving framework may support multiple serving techniques, we first identify, from the official documentation of each framework 
, the APIs associated with the serving methods in our taxonomy. For example, ``from tensorrt\_llm.quantization import ...'' is an example of APIs that we extracted for Low-bit Quantization method of TensorRT\_LLM. We then use a combination of Python regular expressions and Abstract Syntax Tree (AST) analysis~\cite{wikipedia_ast,majidi2022empirical} 
to detect occurrences of these APIs within the collected scripts. Specifically, for each efficient serving method, we first constructed a method-to-API mapping from the official documentation of the selected frameworks, and then used AST analysis to detect relevant imports and parameters in Python scripts, complemented by regular expressions to capture method-related parameter names. By mapping API usages to their corresponding serving methods, we can determine which techniques are adopted in practice and quantify their prevalence across open-source LLM-based software systems.

\subsection{Clustering Repositories}


After identifying the repositories that adopt LLM serving frameworks, we further analyze their repository-level characteristics to understand what types of software systems use these frameworks in practice. To better understand the diversity of LLM serving-related repositories and identify higher-level patterns across their functionalities, we cluster the collected repositories into semantically coherent groups. This enables us to move beyond individual repository-level analysis and instead characterize broader themes, such as deployment-oriented systems, optimization tools, or benchmarking frameworks.

The clustering process consists of two main steps. First, we construct structured textual representations of each repository. Second, we apply a topic-driven clustering pipeline over these representations to identify groups of repositories with similar characteristics.

\subsubsection{Repository Summary Generation}
\label{Repo_summary_Gen}

To construct a consistent representation of each repository, we generate structured summaries based on three complementary metadata sources: the \textit{README}, the \textit{about} section, and repository \textit{keywords}. These sources are selected because they capture different levels of abstraction about a project. The \textit{README} typically provides detailed descriptions of functionality, usage, and design rationale~\cite{github_readme_document,wang2023study}, while the \textit{about} section offers a concise high-level description. In addition, repository \textit{keywords} (or topics) provide compact semantic labels describing the repository’s purpose, domain, or technology stack~\cite{github_topics_document}.

We leverage these complementary signals to generate a unified structured summary for each repository using GPT-4o mini~\cite{openai_chatgpt}. The prompt instructs the model to extract four key dimensions:

\begin{itemize}
    \item \textbf{Repository intent}: the high-level objective of the repository (e.g., serving, training, benchmarking, or deployment).
    \item \textbf{Technical focus}: the main techniques or mechanisms implemented.
    \item \textbf{Primary use case}: the application scenarios supported by the repository.
    \item \textbf{System design}: the architectural or system-level organization of the solution.
\end{itemize}

These dimensions are designed to provide complementary perspectives on each repository. \textit{Repository intent} captures the overall goal, which is often implicit or inconsistently described in raw documentation~\cite{github_readme_document,wang2023study,hellman2021generating}. \textit{Technical focus} distinguishes repositories that share similar goals but rely on different underlying methods, such as quantization, parallelism, or scheduling strategies in LLM serving systems~\cite{miao2025towards,pan2025survey}. \textit{Primary use case} connects the repository to concrete application scenarios such as question answering, code generation, or retrieval-augmented workflows~\cite{pan2025survey}. Finally, \textit{system design} captures how the repository organizes and deploys its components in practice, which is essential for understanding real-world serving architectures~\cite{miao2025towards,pan2025survey,soliman2025mining}.

Overall, these structured summaries provide a normalized and semantically enriched representation of repositories, enabling systematic comparison and downstream clustering.

\subsubsection{Summary Clustering}
\label{Summ_clustering}

We cluster repositories based on the generated summaries in order to identify recurring themes in LLM serving-related software systems. To ensure robustness of the results, we evaluate multiple text preprocessing strategies and compare their impact on clustering quality.

\textbf{Text preprocessing.}
Following prior work~\cite{kumi2024uncovering,jafari2025prioritizing}, we construct three versions of the dataset: (i) raw summaries, (ii) lightly preprocessed summaries, and (iii) fully preprocessed summaries.

All versions first undergo a common initial cleaning stage, which includes removing repeated template phrases (e.g., \textit{repository intent}, \textit{technical focus}, \textit{primary use case}, \textit{system design}), normalizing whitespace, converting text to lowercase, expanding contractions, removing null entries, and eliminating duplicates.

In the \textit{light preprocessing} variant, we tokenize the text using a regex-based tokenizer retaining only alphabetic tokens, remove standard and domain-specific stopwords, and reconstruct the cleaned tokens into text form.

In the \textit{full preprocessing} variant, we further apply part-of-speech filtering (retaining only nouns and verbs), followed by lemmatization, and removal of non-alphabetic or residual stopword tokens.

\textbf{Clustering and topic modeling pipeline.}
We adopt a topic-driven clustering approach based on semantic text embeddings. First, each repository summary is encoded using the SentenceTransformer model \textit{all-MiniLM-L6-v2}~\cite{reimers2019sentence,janssens2025llm}, producing dense semantic embeddings that capture contextual similarity between repositories.

Next, we reduce the dimensionality of these embeddings using UMAP~\cite{kumi2024uncovering,taraghi2026real}, which preserves local semantic structure while improving clustering efficiency. The reduced embeddings are then clustered using HDBSCAN, a density-based clustering algorithm that does not require pre-specifying the number of clusters and can identify outliers as noise~\cite{kumi2024uncovering}.

Finally, we apply BERTopic~\cite{kumi2024uncovering,janssens2025llm} to extract interpretable topic representations for each cluster by identifying the most representative terms within each group. In this setup, HDBSCAN is used as the underlying clustering component of BERTopic, while UMAP and sentence embeddings provide the semantic representation space.

\textbf{Model selection.}
To select the best preprocessing configuration, we evaluate the resulting topic models using two coherence metrics: \(c_v\) and \(u\_mass\). The \(c_v\) metric measures the semantic coherence of top topic words (higher values indicate more coherent and interpretable topics), while \(u\_mass\) captures statistical co-occurrence patterns in the corpus~\cite{kumi2024uncovering} (values closer to zero indicate better coherence). We select the configuration that achieves the best overall balance between these two metrics.

This entire pipeline is applied separately to the three preprocessing variants, and the resulting topic structures are compared to assess robustness of the clustering outcomes.

\section{Result}
\label{results}
This section presents the findings of our study by addressing the research questions introduced earlier.

\subsection{RQ1. What are the most popular and widely adopted LLM serving frameworks and efficient serving methods?}    
\label{rq1}

Our analysis of
LLM serving frameworks shows that they provide a diverse set of serving methods and functionalities, while attracting varying levels of attention from developers. Furthermore, the results indicate that developers tend to favor certain LLM serving methods over others.

\subsubsection{What are the most popular LLM serving frameworks?} 
\label{rq1.1}

To identify the most popular LLM serving frameworks, we collect detailed information from the GitHub repositories associated with each tool identified in Section~\ref{subsec:extract_serv_tool}. Following prior studies~\cite{schreiber2025security,das2022empirical}, we use GitHub repository popularity metrics (the number of stars and forks) to rank these tools.
We then manually examine all the collected frameworks and filter them based on the following exclusion criteria:

\begin{itemize}
    \item Frameworks that are not specifically designed for LLM serving but instead provide general-purpose distributed computing infrastructures (e.g., Ray~\cite{moritz2018ray}) without implementing LLM-specific serving optimizations.
    \item Frameworks that do not provide a clear description or documentation of the cost-efficient LLM serving methods they implement (e.g., OpenLLM~\cite{OpenLLM_document}).
    \item Frameworks that primarily target proprietary or closed-source LLMs (e.g., RouteLLM~\cite{RouteLLM_document}), limiting their applicability and reproducibility within the scope of our study.
\end{itemize}

After applying these criteria, \textbf{we identified five representative and widely used LLM serving frameworks:
vLLM~\cite{vllm_document}, SGLang~\cite{sglang_document}, TensorRT-LLM~\cite{TensorRT_LLM_document}, LMDeploy~\cite{LMDeploy_document}, and FlashInfer~\cite{FlashInfer_document}}.
Table~\ref{tab:rq1.1} presents the ten most popular frameworks based on GitHub stars and forks as of November 2025. The upper portion of the table contains the frameworks retained for further analysis, whereas the lower portion lists those excluded during the filtering process. The remainder of this paper focuses on the five selected frameworks as a representative set of open-source LLM-serving solutions.


\begin{table}[]
\centering
\caption{Top 10 LLM serving frameworks ranked by GitHub stars, with the five selected frameworks for subsequent RQs }
\label{tab:rq1.1}
\begin{tabular}{l p{3cm} l l}
\hline
& \textbf{Framework} & \textbf{No. of stars} & \textbf{No. of Forks}\\
\hline
\multirow{5}{*}{\rotatebox[origin=c]{90}{Selected}}& vllm \cite{vllm_document} & 56.3K & 9.6K\\
& SGLang \cite{sglang_document} & 17.2K & 2.7K\\
& TensorRT-LLM \cite{TensorRT_LLM_document} & 13.2K & 2.2K\\
& LMDeploy \cite{LMDeploy_document} & 6.9K & 599\\

& FlashInfer \cite{FlashInfer_document} & 3.6K & 459\\

\hline

\multirow{5}{*}{\rotatebox[origin=c]{90}{Filtered out}} & Ray \cite{ray_document} & 38.6K & 6.7\\
& OpenLLM \cite{OpenLLM_document} & 12.3K & 805\\
& RouteLLM \cite{RouteLLM_document} & 4.2K & 330\\
& LighLLM \cite{LightLLM_document} & 3.5K & 276\\
& FastDeploy \cite{FastDeploy_document} & 3.4K & 604\\

\hline
\end{tabular}

\end{table}

\subsubsection{What are the most adopted LLM serving frameworks?}
\label{rq1.2}
To assess framework adoption, we analyzed the extracted serving scripts and associated repositories that invoke each framework, as described in Section~\ref{sec:method_extract_repos}. Table~\ref{tab:rq1.2} reports the number of unique repositories using each framework before and after filtering out unpopular, inactive, personal, and immature repositories according to the criteria defined in Section~\ref{sec:method_extract_repos}.

\textbf{The results indicate that vLLM is by far the most widely adopted framework, with 1,821 repositories remaining after filtering}. Its dominant adoption may be partially explained by its early emergence as one of the first frameworks specifically designed to improve the efficiency and cost-effectiveness of LLM serving.

\textbf{Interestingly, popularity and adoption do not always align. For example, FlashInfer ranks last among the selected frameworks in terms of GitHub popularity, yet it is the third most adopted framework}. One possible explanation is its relatively recent release, which has limited the time available to accumulate stars and forks despite growing practical usage.

Conversely, TensorRT-LLM ranks third in popularity but exhibits the lowest adoption among the selected frameworks. Although it is widely recognized as a high-performance serving solution, its reliance on specialized NVIDIA hardware~\cite{TensorRT_LLM_document} may restrict its accessibility and consequently limit its adoption across open-source repositories which requires further investigation.

\textbf{Overall, these findings suggest that popularity metrics alone do not necessarily reflect real-world adoption. Factors such as hardware requirements, ease of integration, maturity, and ecosystem support appear to play an important role in determining the extent to which a serving framework is adopted in practice.}


\begin{table}[]
\centering
\caption{Top five LLM serving frameworks ranked by the number of repositories adopting each framework.}
\label{tab:rq1.2}
\begin{tabular}{p{3cm} l l}
\hline
\textbf{framework} & \textbf{Repos (Before Filtering)} & \textbf{Repos (After Filtering)} \\
\hline
vLLM & 7,057 & 1,821 \\
SGLang & 315 & 54\\
FlashInfer & 371 & 52\\
LMDeploy & 463 & 45 \\
TensorRT-LLM & 70 & 15 \\
\hline
\end{tabular}

\end{table}

\subsubsection{What are the most used efficient LLM serving methods?}
\label{rq1.3}

To identify the most widely used LLM serving methods, we first extracted the serving techniques supported by the five selected frameworks following the procedure described in Section~\ref{subsec:serv_method}. Table~\ref{tab:coveredmethods} summarizes the serving methods implemented by each framework and their adoption across repositories.

When mapping framework-specific methods to the taxonomy presented in Fig.~\ref{fig:servingmethods}, we observed that several methods could be assigned to existing subcategories but did not fit any of the predefined 
sub-subcategories mentioned in Figure \ref{fig:servingmethods}. To avoid forcing these methods into wrong classes
, we introduced three additional sub-subcategories: \textit{Additional Decoding Algorithms}, \textit{Additional Parallel Computation Methods}, and \textit{Additional Kernel Optimization Methods}. Specifically, \textit{Additional Parallel Computation Methods} include \textit{Tensor Parallelism}, \textit{Pipeline Parallelism}, \textit{Expert Parallelism}, \textit{Data Parallelism}, and \textit{Model Parallelism}; \textit{Additional Decoding Algorithms} include \textit{Guided Decoding}; and \textit{Additional Kernel Optimization Methods} include \textit{CUDA/HIP Graph Execution}, \textit{FlashAttention-based Execution}, and \textit{High-Performance CUDA Kernels}.

To assess method adoption, we searched the extracted Python scripts for APIs associated with each serving method, following the approach described in Section~\ref{sec:method_extract_repos}. 
Table \ref{tab:coveredmethods} reports both the number and percentage of repositories using each method. The total number of repositories associated with each framework is shown in the table header, while the two most frequently used methods within each framework are highlighted in bold.

\begin{table}[!htbp]
\centering
\caption{Number and percentage of unique repositories using each LLM serving methods. For each framework, \textit{Method Support} indicates whether a method is supported or implemented by the framework, \textit{No. of repo} reports the number of unique repositories using that method, and \textit{Ratio\%} reports the percentage relative to all unique repositories using that framework. The numbers in parentheses beside each framework name denote the total number of unique repositories that used that framework. \textit{Repos} means repositories. The two most commonly used serving methods for each framework are shown in bold text.}
\label{tab:coveredmethods}
\scriptsize
\renewcommand{\arraystretch}{1.4}
\resizebox{\textwidth}{!}{%
\begin{tabular}{|p{3cm}!{\vrule width 1.2pt}
>{\centering\arraybackslash}p{0.8cm}|>{\centering\arraybackslash}p{1cm}!
{\vrule width 1.2pt}
>{\centering\arraybackslash}p{0.8cm}|>{\centering\arraybackslash}p{1cm}!
{\vrule width 1.2pt}
>{\centering\arraybackslash}p{0.8cm}|>{\centering\arraybackslash}p{1cm}!
{\vrule width 1.2pt}
>{\centering\arraybackslash}p{0.8cm}|>{\centering\arraybackslash}p{1cm}!
{\vrule width 1.2pt}
>{\centering\arraybackslash}p{0.8cm}|>{\centering\arraybackslash}p{1cm}|>
{\centering\arraybackslash}p{1cm}|
}
\hline

\multicolumn{1}{|p{3cm}!{\vrule width 1.2pt}}{\multirow{2}{*}{\textbf{Method}}}
& \multicolumn{2}{c!{\vrule width 1.2pt}}{\textbf{vLLM (1,821 Repos)}}
& \multicolumn{2}{c!{\vrule width 1.2pt}}{\textbf{SGLang (54 Repos)}}
& \multicolumn{2}{c!{\vrule width 1.2pt}}{\textbf{TensorRT-LLM (15 Repos)}}
& \multicolumn{2}{c!{\vrule width 1.2pt}}{\textbf{LMDeploy (45 Repos)}}
& \multicolumn{2}{c!{\vrule width 1.2pt}}{\textbf{FlashInfer (52 Repos)}}
& \multicolumn{1}{c|}{\textbf{Total}}
\\ \cline{2-12}


& \makecell{\textbf{Method}\\\textbf{Support}} & \makecell{\textbf{\# Repos}\\\textbf{(Ratio \%)}} 
& \makecell{\textbf{Method}\\\textbf{Support}} & \makecell{\textbf{\# Repos}\\\textbf{(Ratio \%)}} 
& \makecell{\textbf{Method}\\\textbf{Support}} & \makecell{\textbf{\# Repos}\\\textbf{(Ratio \%)}} 
& \makecell{\textbf{Method}\\\textbf{Support}} & \makecell{\textbf{\# Repos}\\\textbf{(Ratio \%)}} 
& \makecell{\textbf{Method}\\\textbf{Support}} & \makecell{\textbf{\# Repos}\\\textbf{(Ratio \%)}}
& \makecell{\textbf{\# Repo}}\\ 
\hline

Non-autoregressive Decoding &-&-&-&-&-&-&-&-&-&-&- \\ \hline

Speculative Decoding & \greencirclecheck &9 (0.49\%) & \greencirclecheck & 1 (1.85\%)&  \greencirclecheck & 1 (6.6\%)&-&-&-&- & 11 \\ \hline

Early Exiting & \greencirclecheck &1 (0.05\%) &-&-&-&-&-&-&-&-& \cellcolor{red!10}  1\\ \hline

Cascade Inference & \greencirclecheck &-&-&-&-&-&-&-&  \greencirclecheck & 12 (23\%) & 12\\ \hline

Additional Decoding Algorithms &-&- &-&-& \greencirclecheck & 1 (6.6\%)&-&-&-&- & \cellcolor{red!10}  1\\ \hline

Config Downsizing &-&-&-&-&-&-&-&-&-&-&-\\ \hline

Attention Simplification &-&-&-&-&-&-&-&-&-\greencirclecheck & 1 (1.9\%) & \cellcolor{red!10}  1\\ \hline

Recurrent Unit &-&-&-&-&-&-&-&-&-&-&-\\ \hline

Activation Sharing &-&-&-&-&-&-&-&-&-&-&-\\ \hline

Conditional Computing &-&-&-&-&-&-&-&-& \greencirclecheck &10 ( 19.2\%) & 10\\ \hline


Knowledge Distillation &-&-&-&-&-&-&-&-&-&-&-\\ \hline

Network Pruning & \greencirclecheck & 202 (11\%)&-&-& \greencirclecheck & 1 (6.6\%)&-&-&-&-& \cellcolor{blue!15}  \textbf{203}\\ \hline


Low-bit Quantization methods & \greencirclecheck & 140 (7\%)& \greencirclecheck &4 (7.4\%) & \greencirclecheck & 2 (13.3\%)& \greencirclecheck & & \greencirclecheck &15 (28\%) & 161\\ \hline

Cloud Scaling &-&-&-&-&-&-&-&-&-&-&-\\ \hline

Decentralized Inference &-&-&-&-&-&-&-&-&-&-&-\\ \hline

Additional Parallel Computation methods & \greencirclecheck & \textbf{977} (\textbf{53\%}) & \greencirclecheck & \textbf{22} (\textbf{40\%})& \greencirclecheck &3 (20\%)& \greencirclecheck &-& \greencirclecheck &8 (15\%) & \cellcolor{blue!15} \textbf{1010}\\ \hline

Memory Management methods & \greencirclecheck & \textbf{396} (\textbf{21\%}) & \greencirclecheck & \textbf{6} (\textbf{11\%})&  \greencirclecheck &\textbf{8} (\textbf{53\%})& \greencirclecheck &\textbf{7} (\textbf{15\%})& \greencirclecheck &\textbf{34} (\textbf{65\%}) & \cellcolor{blue!15} \textbf{451}\\ \hline

Kernel Fusion & \greencirclecheck &-&- &-&- &-&- &-&  \greencirclecheck & \textbf{46} (\textbf{88\%})& 46\\ \hline

Tailored Attention &-&-&-&-&-&-&-&-& \greencirclecheck &27 (51\%) & 27\\ \hline

Variable Sequence Length &-&-&-&-&-&-&-&-&-&-&-\\ \hline

Automatic Compilation &-&-&-&-&-&-&-&-&-&-&-\\ \hline

Additional Kernel Optimizations methods & \greencirclecheck &22 (1.2\%) &- &-&-&- & \greencirclecheck &-& \greencirclecheck & 2 (3.8\%) & 24\\ \hline

Request Scheduling methods & \greencirclecheck & 121 (6.6\%)& \greencirclecheck &1 (1.85\%)& \greencirclecheck &\textbf{7} (\textbf{46\%})&  \greencirclecheck &-&-&-& 129\\ 
\hline

\end{tabular}%
}

\end{table}

\textbf{The results reveal that \textit{Additional Parallel Computation Methods} (1,010 repositories), \textit{Memory Management Methods} (451 repositories), and \textit{Network Pruning} (203 repositories) are the most widely adopted serving-method categories overall. These findings suggest that developers primarily focus on addressing the two dominant bottlenecks of LLM serving: computational scalability and memory efficiency.}

However, adoption patterns vary considerably across frameworks. For example, repositories using FlashInfer most frequently employ \textit{Kernel Fusion}, even though the framework also supports parallel computation and memory-management techniques. This observation is consistent with FlashInfer's design objective of providing highly optimized GPU kernels for inference acceleration~~\cite{ye2025flashinfer}. Consequently, \textbf{developers appear to adopt FlashInfer primarily for its kernel-level optimizations rather than its broader serving capabilities.}

The prominence of parallel computation and memory-management techniques also helps explain the widespread adoption of vLLM. Prior studies identify computational parallelism and memory consumption as the primary performance bottlenecks in LLM serving~\cite{kwon2023efficient,qianli2025mell,chen2026universal}. vLLM was explicitly designed to address these challenges through efficient batching mechanisms and KV-cache management~\cite{vllm_document}. By directly targeting the most critical efficiency concerns, vLLM provides capabilities that align closely with practitioners' needs, which likely contributes to its dominant adoption.

In contrast, \textbf{methods such as early exiting, speculative decoding, attention simplification, and other architecture-level optimizations exhibit very limited adoption.} These techniques typically require modifications to model architectures or inference procedures and often necessitate retraining or fine-tuning~\cite{elhoushi2024layerskip,li2024llm}. Moreover, many of these approaches remain primarily research-oriented and have not yet been widely integrated into production-ready serving frameworks~\cite{miao2025towards}. Their adoption may also be hindered by compatibility constraints with pre-trained models and potential trade-offs between efficiency gains and output quality~\cite{pope2023efficiently,shi2025systematic}. Consequently, \textbf{practitioners appear to favor optimization techniques that are easier to deploy and better supported by existing serving infrastructures.}

Interestingly, some serving methods are implemented in existing frameworks but are rarely or never observed in practice. For example, although LMDeploy supports LoRA-based serving, we found no evidence of its use in the analyzed repositories. One possible explanation is that LoRA primarily benefits adapter-based fine-tuning scenarios~~\cite{dettmers2023qlora}, whereas many applications deploy pre-trained models without additional adaptation. Furthermore, LoRA serving introduces additional operational complexity, including the management of base models and adapter weights, as well as compatibility considerations~\cite{hu2022lora}. As a result, LoRA-based serving remains relatively uncommon in production deployments~\cite{mao2025survey}.

Finally, \textbf{several serving methods identified in the literature; including non-autoregressive decoding, configuration downsizing, recurrent architectures, and activation sharing, are not currently implemented in any of the studied frameworks.} This absence likely reflects practical deployment considerations. Many of these methods require substantial departures from the transformer-based autoregressive paradigm that underpins most modern LLMs~~\cite{sun2023retentive,peng2023rwkv}. Integrating such approaches would therefore require significant changes to both model architectures and serving infrastructures~\cite{gu2017non}. In addition, these techniques are often evaluated under specialized research settings and may depend on customized training procedures, limiting their applicability to existing pre-trained models~\cite{miao2025towards}. In contrast, serving frameworks tend to prioritize general-purpose optimizations, such as parallel computation and memory management, that can be broadly applied across models and workloads while preserving output quality and compatibility.

\begin{tcolorbox}[breakable]
\textbf{Finding 1.} Among the selected LLM-specific serving frameworks, vLLM is the most visible framework in terms of GitHub popularity and the most adopted framework in the filtered repositories. Across the studied repositories, Additional Parallel Computation, Memory Management, and Network Pruning are the most frequently observed serving-method categories, while method adoption remains framework-specific, with repositories that use FlashInfer showing a strong concentration around Kernel Fusion.
\end{tcolorbox}


\subsection{RQ2. What are the most used combinations of LLM serving frameworks, efficient serving methods, and LLMs?}
\label{rq2}

This section examines the extent to which developers combine LLM serving methods, either within a single serving framework or across multiple frameworks, to improve LLM serving efficiency. Our analysis of repositories that use LLM serving frameworks reveals that developers frequently employ multiple serving methods provided by the same framework and, in some cases, integrate several frameworks and their associated optimization techniques within a single system. These combinations are used to address different performance bottlenecks and enhance serving efficiency. Furthermore, this section identifies the most common combinations of serving methods and highlights the LLMs most frequently deployed with each serving framework.

\subsubsection{What are the most used combinations of efficient LLM serving methods?} 
\label{rq2.1}
\begin{figure*}[!htbp]
\centering

\begin{tabular}{cc}

\subfloat[vLLM\label{fig:rq2.2_vllm}]{%
    \includegraphics[width=0.47\textwidth]{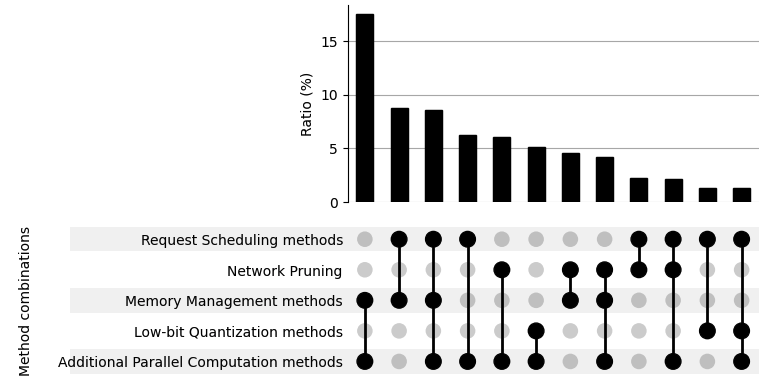}
}
&
\subfloat[SGLang\label{fig:rq2.2_sglang}]{%
    \includegraphics[width=0.47\textwidth]{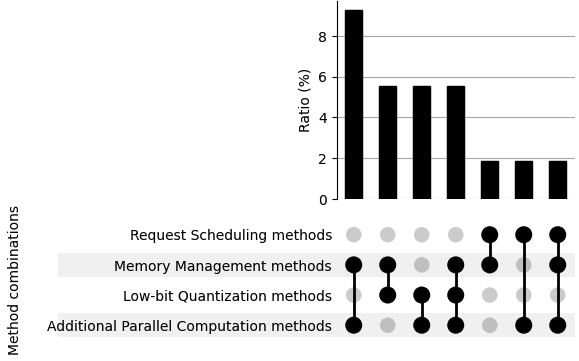}
}
\\[0.3cm]

\subfloat[TensorRT-LLM\label{fig:rq2.2_tensorrt}]{%
    \includegraphics[width=0.47\textwidth]{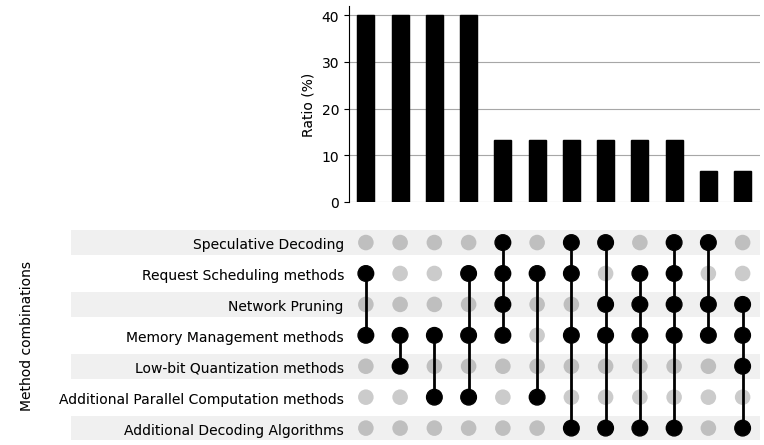}
}
&
\subfloat[FlashInfer\label{fig:rq2.2_flashinfer}]{%
    \includegraphics[width=0.47\textwidth]{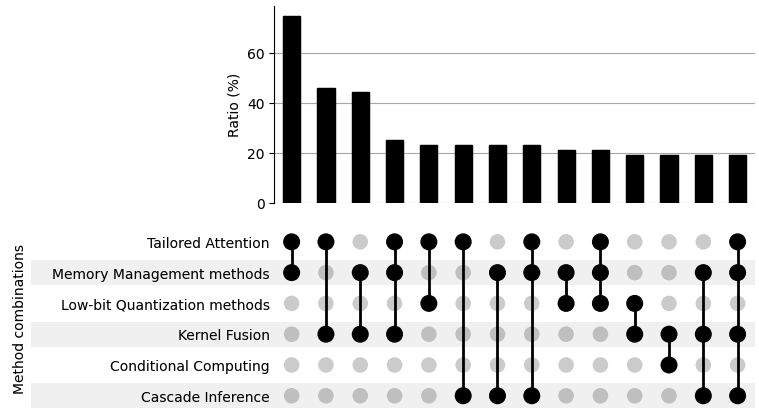}
}

\end{tabular}

\caption{UpSet plot of the most frequent method combinations of each of the vLLM, SGLang, TensorRT-LLM, and FlashInfer frameworks, where bars show usage ratio (\%) w.r.t. the total number of repositories using each framework and connected dots indicate the methods in each combination.}
\label{fig:rq2.2_all}
\end{figure*}

In this section, we analyze how serving methods are combined within each studied framework. Specifically, we examine the combinations of methods that are most frequently used together to improve LLM serving efficiency. Figures \ref{fig:rq2.2_vllm}, \ref{fig:rq2.2_sglang}, \ref{fig:rq2.2_tensorrt}, \ref{fig:rq2.2_flashinfer} present the most common method combinations observed in vLLM, SGLang, TensorRT-LLM, and FlashInfer, respectively.

LMDeploy is not included in these figures because, among the repositories analyzed, only Memory Management methods were used, and no combinations involving other LMDeploy-supported methods were observed. Consequently, the absence of an LMDeploy plot reflects the usage patterns found in the studied repositories rather than any limitation of LMDeploy’s documented capabilities.  

The patterns shown in Figure \ref{fig:rq2.2_all} provide insights into how LLM serving methods are combined in practice across open-source repositories. Rather than indicating which combinations are the most effective, the results reveal which combinations are most commonly adopted by developers. Consequently, the observed combinations should be interpreted as usage patterns rather than as evidence of superior serving performance.

These findings may help developers of LLM serving frameworks identify combinations of methods that are frequently used together in practice. Framework developers can use this information to improve documentation, APIs, tutorials, and deployment examples by considering commonly adopted combinations in addition to describing individual optimization methods in isolation.

The results may also assist developers of LLM-based applications during framework configuration and deployment. Instead of selecting serving optimizations independently, practitioners can use the observed combinations as reference points when exploring configuration choices. For example, \textbf{repositories using vLLM and SGLang frequently combine Memory Management methods with Additional Parallel Computation methods, whereas FlashInfer repositories more often combine Tailored Attention, Memory Management, and Kernel Fusion methods.} These observations do not imply that such combinations are optimal; rather, they indicate that these combinations have been adopted by developers in existing repositories and may therefore warrant further evaluation in new deployments.

More broadly, the findings highlight the importance of studying interactions among serving methods. While prior work often evaluates individual optimization techniques in isolation, \textbf{our results suggest that real-world deployments frequently combine multiple techniques simultaneously.} Understanding how these methods interact, complement one another, or potentially introduce trade-offs remains an open research question.

Finally, the long-tail distribution of observed combinations may be equally informative. Frequently occurring combinations reveal common deployment practices, whereas rarely observed combinations may indicate underexplored design spaces, limited framework support, higher implementation complexity, or simply a lack of adoption within the repositories studied. Future work could investigate the reasons behind both frequent and infrequent combinations and empirically evaluate their impact on serving efficiency, scalability, and resource utilization.

A possible interpretation of the combinations in the Figure \ref{fig:rq2.2_all} is that \textbf{some developers combine serving methods that address different bottlenecks in LLM inference.} \textbf{For vLLM, the combination of Memory Management methods with Additional Parallel Computation methods suggests that repositories may use memory-efficient KV-cache management together with distributed execution to support larger models and improve serving capacity under GPU memory constraints.} This interpretation could be supported by the vLLM paper, which states that vLLM uses ``block-level memory management and preemptive request scheduling'' together with PagedAttention, and also supports models whose sizes exceed the memory capacity of a single GPU \cite{kwon2023efficient}. 
It could also be supported by the vLLM documentation, which states that vLLM supports distributed tensor-parallel and pipeline-parallel inference and serving \cite{vllmDocsParallelism}. 
Therefore, this combination can be interpreted as bringing together two complementary capabilities: reducing KV-cache memory waste and scaling inference across GPU resources \cite{kwon2023efficient,vllmDocsParallelism}.
The combinations involving Request Scheduling methods further suggest that vLLM users may try to improve the latency-throughput trade-off. For example, vLLM documentation explains that chunked prefill can improve inter-token latency by prioritizing decode requests and can improve GPU utilization by batching compute-bound prefill requests with memory-bound decode requests \cite{vllmDocsChunkedPrefill}. 
Thus, the three-way combination of Request Scheduling, Memory Management, and Additional Parallel Computation could be interpreted as a possible practical configuration for jointly managing GPU memory, scheduling mixed prefill/decode workloads, and scaling execution across GPU resources \cite{kwon2023efficient,vllmDocsParallelism,vllmDocsChunkedPrefill}.

For SGLang, the combination of Memory Management methods with Additional Parallel Computation methods could be interpreted through the interaction between RadixAttention, batching, and parallel execution. The SGLang paper \cite{zheng2024sglang} explains that RadixAttention enables KV-cache reuse by retaining the cache in a radix tree, and that this design supports efficient prefix search, reuse, insertion, and eviction. The same paper also states that RadixAttention is compatible with continuous batching, paged attention, and tensor parallelism \cite{zheng2024sglang}. Therefore, when repositories combine Memory Management methods with Additional Parallel Computation methods, the combination can be interpreted as bringing together prefix/KV-cache reuse and parallel execution. This may help repositories reduce redundant computation and memory usage for requests with shared prompt prefixes, while still allowing the system to use tensor parallelism for larger models on multiple GPUs \cite{zheng2024sglang}. The additional appearance of Low-bit Quantization methods suggests a complementary optimization path: quantization can reduce the computational and memory costs of inference by representing weights and activations with lower-precision data types \cite{huggingfaceQuantization}.

\textbf{For TensorRT-LLM, Figure \ref{fig:rq2.2_tensorrt} shows that Memory Management methods appear in 11 of the 12 observed method combinations. 
This may suggest that memory-related optimization is a recurring part of the observed TensorRT-LLM combinations.} TensorRT-LLM lists In-Flight Batching and Paged Attention together with KV Cache Management as advanced optimization features~\cite{tensorrtLlmOverview}.
In-flight batching improves GPU utilization and reduces latency, while paged KV cache manages the KV cache using blocks assigned to different requests \cite{tensorrtLlmOverview,tensorrtLlmGptAttention}. 
Therefore, the combination of Memory Management methods with Request Scheduling methods could be interpreted as bringing together request handling and cache management. Combinations with Additional Parallel Computation methods may also be useful when a model cannot fit in a single GPU's memory or when a single GPU cannot provide the desired performance, since these are cases where multi-GPU parallelism becomes necessary \cite{tensorrtLlmParallelStrategy}.
Finally, combinations with Low-bit Quantization methods suggest an additional efficiency direction, because quantization reduces memory footprint and computational cost \cite{tensorrtLlmQuantization}. Overall, these \textbf{combinations could be bring together memory management, scheduling, parallel execution, and reduced precision.}

\textbf{For FlashInfer, Figure \ref{fig:rq2.2_flashinfer} shows that the observed combinations are concentrated around Tailored Attention, Memory Management, and Kernel Fusion. This pattern can be interpreted as an emphasis on attention- and kernel-level efficiency, since efficient GPU attention kernels are important for high-throughput and low-latency LLM inference serving \cite{ye2025flashinfer}.} The combination of Tailored Attention with Memory Management can be interpreted as bringing together attention execution and memory-efficient KV-cache handling, since FlashInfer addresses KV-cache storage heterogeneity using block-sparse and composable formats to improve memory access and reduce redundancy \cite{ye2025flashinfer}. The combination of Tailored Attention with Kernel Fusion can also be interpreted as targeting latency reduction, since the FlashInfer paper reports latency reduction in long-context inference settings \cite{ye2025flashinfer}. 

The findings of Figure \ref{fig:rq2.2_all} can help developers of LLM serving frameworks understand what efficient seving methods are combined in practice by other repositories and developers.
\textbf{The findings of Figure \ref{fig:rq2.2_all} will help framework developers improve APIs, documentation, and examples for commonly used method combinations along with documenting each method as an isolated feature.}
\textbf{The results can also help developers of LLM-based applications choose and configure serving frameworks more effectively by trying the common combinations of methods and check whether they perform better regarding efficiency.} Instead of selecting serving methods independently, developers can use these figures to identify combinations already used in existing GitHub repositories.
For example, \textbf{vLLM and SGLang users may focus on combinations involving Memory Management and Parallel Computation, while FlashInfer users may focus on combinations involving Tailored Attention, Memory Management, and Kernel Fusion to check whether the efficiency improves.}
The findings can also guide researchers in LLM systems and software engineering. \textbf{The results show which serving-method combinations are common in practice and which combinations are less visible in open-source repositories. This can motivate future research on how serving methods interact when they are used together, rather than studying each method only in isolation.}

\subsubsection{What are the most used combinations of LLM serving frameworks?}
\label{rq2.2}
In this section, we examine whether different LLM serving frameworks are used together within the same GitHub repository. While Section \ref{rq2.1} investigated whether multiple serving methods of a single framework are used in combination to further improve the efficiency of LLM serving, it remains unclear whether different frameworks—and their corresponding method categories—are also combined in practice. Therefore, this section explores whether GitHub repositories 
use multiple LLM serving frameworks together and, if so, which methods from these frameworks are used within the same repository.

Table~\ref{tab:framework-combination-ratio} reports the count and percentage of repositories that combine each  serving framework with other frameworks, showing that framework co-usage differs substantially across the studied frameworks.
For example, vLLM has the largest number of repositories that combine it with at least one other framework, with 28 repositories. However, because vLLM is also by far the most frequently used framework in the dataset, this represents only 1.54\% of all repositories using vLLM, which is the lowest co-usage ratio in the table.
This suggests that \textbf{although vLLM is widely adopted, most vLLM repositories use it without combining it with another studied serving framework.} One possible reason could be that vLLM is designed as a high-throughput distributed LLM serving engine based on PagedAttention, block-level KV-cache memory management, and preemptive request scheduling addressing the main bottleneck of efficient LLM serving, such as high memory usage. Therefore, it could be used as an standalone serving engine \cite{kwon2023efficient}.
In contrast, \textbf{FlashInfer has the highest co-usage ratio, with 23 out of 52 repositories, or 44.23\%, also using at least one other studied serving framework A possible interpretation is that FlashInfer is often adopted as a specialized attention- and kernel-level component within a broader LLM serving stack.} Also, Ye et al. describe FlashInfer as an efficient and customizable attention engine for LLM serving and report its integration with other serving frameworks \cite{ye2025flashinfer}. Since FlashInfer also supports KV-cache storage heterogeneity, customizable attention templates, and dynamic scheduling \cite{ye2025flashinfer}, its higher co-usage ratio may reflect its role as a specialized optimization component that can complement broader serving frameworks.


\begin{table}[!htbp]
\centering
\small
\setlength{\tabcolsep}{6pt}
\renewcommand{\arraystretch}{1.25}
\caption{The number and the ratio of repositories that use a combination of LLM serving frameworks. \textit{`no of repos'} represents the total number of repositories that adopted each serving frameworks. \textit{`Num of repos using combined frameworks'} also indicates the number of repositories that combine related LLM serving framework with other frameworks.}
\label{tab:framework-combination-ratio}
\begin{tabular}{|l|c|}
\hline
\textbf{Framework (num of repos)} & \textbf{Num of repos using combined frameworks} \\
\hline
vLLM (1,821) & 28 (1.54\%) \\
\hline
SGLang (54) & 11 (20.37\%) \\
\hline
TensorRT-LLM (15) & 4 (26.67\%) \\
\hline
LMDeploy (45) & 2 (4.44\%) \\
\hline
FlashInfer (52) & 23 (44.23\%) \\
\hline
\end{tabular}
\end{table}

Although Table \ref{tab:framework-combination-ratio} provides an overview of the extent to which each LLM serving framework is used in combination with other frameworks, it does not explain which framework combinations occur or which serving methods are used in those multi-framework repositories. Therefore, we conducted a more detailed analysis on the repositories that use multiple LLM serving frameworks. Table \ref{tab:framework-method-category-combinations} summarizes the framework combinations observed in these repositories. To provide a more fine-grained view, Table \ref{tab:repo-framework-methods-cleaned} presents the repository-level results for all repositories that use multiple LLM serving frameworks. Each row in Table \ref{tab:repo-framework-methods-cleaned} corresponds to one repository. For each repository, the table reports the frameworks used together, the detected APIs or parameters related to serving methods in the “APIs or parameters used” column, and the mapped categories of those APIs or parameters in the “Method category used” column. To focus on recurring framework co-usage patterns, Tables~\ref{tab:framework-combination-ratio} and \ref{tab:framework-method-category-combinations} reports only the framework combinations that appear in at least four repositories. Combinations with fewer than four occurrences provide limited evidence for drawing meaningful conclusions about common co-usage practices, as they may reflect repository-specific design decisions rather than recurring patterns across projects. This threshold helps keep the analysis focused on combinations with stronger empirical support and reduces the risk of over-interpreting sparse observations.

\begin{table}[!htbp]
\centering
\scriptsize
\setlength{\tabcolsep}{3pt}
\renewcommand{\arraystretch}{1.25}
\caption{Co-occurrence of LLM serving frameworks and method categories in multi-framework repositories. Each row corresponds to one framework combination. The “Framework combination” column lists the frameworks used together. The “Methods category used from each framework” column reports the method categories observed for each framework in that combination and the “\# Repos” column reports the number of repositories in which each combination appears. Bold method names indicate method categories implemented by one framework but not by the other frameworks in the same combination.}
\label{tab:framework-method-category-combinations}
\begin{tabularx}{\textwidth}{
    |>{\raggedright\arraybackslash}p{3.1cm}
    |>{\raggedright\arraybackslash}X
    |>{\centering\arraybackslash}p{0.8cm}|
}
\hline
Framework combination & Method Category used from each framework & \# Repos \\
\hline

FlashInfer (34.62\%), vLLM (0.99\%) & 
FlashInfer: \textbf{Tailored Attention}, \textbf{Cascade Inference}, Additional Parallel Computation methods, \textbf{Kernel Fusion}, \textbf{Conditional Computing}, Memory Management methods, Low-bit Quantization methods, \textbf{Attention Simplification}

vLLM: Additional Parallel Computation methods, \textbf{Request Scheduling methods}, Memory Management methods, Low-bit Quantization methods, \textbf{Speculative Decoding methods}, \textbf{Network Pruning}
& 18 \\
\hline

SGLang (9.26\%), vLLM (0.27\%) & 
SGLang: Memory Management methods, Additional Parallel Computation methods

vLLM: \textbf{Network Pruning}, Additional Parallel Computation methods
& 5 \\
\hline

FlashInfer (7.69\%), SGLang (7.41\%) & 
FlashInfer: \textbf{Tailored Attention}, \textbf{Cascade Inference}, Additional Parallel Computation methods, \textbf{Kernel Fusion}, \textbf{Conditional Computing}, Memory Management methods, Low-bit Quantization methods

SGLang: \textbf{Request Scheduling methods}, Low-bit Quantization methods, Additional Parallel Computation methods, Memory Management methods
& 4 \\
\hline

\end{tabularx}
\end{table}

{\scriptsize
\setlength{\tabcolsep}{2pt}
\renewcommand{\arraystretch}{1.2}

\setlength{\LTleft}{0pt}
\setlength{\LTright}{0pt}
\setlength{\LTcapwidth}{\textwidth}

\begin{longtable}{
    |>{\raggedright\arraybackslash}p{0.16\textwidth}
    |>{\raggedright\arraybackslash}p{0.34\textwidth}
    |>{\raggedright\arraybackslash}
      p{\dimexpr0.50\textwidth-6\tabcolsep-4\arrayrulewidth\relax}|
}
\caption{Repository-level co-usage of LLM serving frameworks, where each row represents one repository and shows the APIs or parameters used for each framework combination and their corresponding mapped method categories.}
\label{tab:repo-framework-methods-cleaned}\\
\hline
\textbf{Framework combination} & \textbf{APIs or parameters used} & \textbf{Method category used} \\
\hline
\endfirsthead

\hline
\textbf{Framework combination} & \textbf{APIs or parameters used} & \textbf{Method category used} \\
\hline
\endhead

FlashInfer, vLLM & \textbf{FlashInfer:} Continuous batching, FlashInfer Attention Kernels, cascade, comm, flashinfer.gemm, fused\_moe, norm, page, quantization, sampling, topk\newline \textbf{vLLM:} Chunked prefill, KVCaching, Multi-LoRA, Quantization, Speculative decoding, parallelism & \textbf{FlashInfer:} Tailored Attention, Cascade Inference, Parallel Computation methods, Kernel Fusion, Conditional Computing, Memory Management methods, Low-bit Quantization methods\newline \textbf{vLLM:} Request Scheduling methods, Memory Management methods, Network Pruning, Quantization, Speculative Decoding, Parallel Computation methods \\
\hline

FlashInfer, vLLM & \textbf{FlashInfer:} Continuous batching, FlashInfer Attention Kernels, cascade, comm, flashinfer.gemm, fused\_moe, norm, page, quantization, sampling, topk\newline \textbf{vLLM:} Chunked prefill, KVCaching, Multi-LoRA, Quantization, parallelism & \textbf{FlashInfer:} Tailored Attention, Cascade Inference, Parallel Computation methods, Kernel Fusion, Conditional Computing, Memory Management methods, Low-bit Quantization methods\newline \textbf{vLLM:} Request Scheduling methods, Memory Management methods, Network Pruning, Quantization, Parallel Computation methods \\
\hline

FlashInfer, vLLM & \textbf{FlashInfer:} FlashInfer Attention Kernels, norm, page, sampling\newline \textbf{vLLM:} CUDA, KVCaching, Multi-LoRA, Quantization, parallelism & \textbf{FlashInfer:} Tailored Attention, Kernel Fusion, Memory Management methods\newline \textbf{vLLM:} Kernel Optimizations methods, Memory Management methods, Network Pruning, Quantization, Parallel Computation methods \\
\hline

FlashInfer, vLLM & \textbf{FlashInfer:} FlashInfer Attention Kernels, logits\_processor, norm, quantization, sampling\newline \textbf{vLLM:} Quantization, Speculative decoding, parallelism & \textbf{FlashInfer:} Tailored Attention, Kernel Fusion, Low-bit Quantization methods\newline \textbf{vLLM:} Quantization, Speculative Decoding, Parallel Computation methods \\
\hline

FlashInfer, vLLM & \textbf{FlashInfer:} FlashInfer Attention Kernels, norm, page, sampling\newline \textbf{vLLM:} CUDA, KVCaching, Multi-LoRA, Quantization, parallelism & \textbf{FlashInfer:} Tailored Attention, Kernel Fusion, Memory Management methods\newline \textbf{vLLM:} Kernel Optimizations methods, Memory Management methods, Network Pruning, Quantization, Parallel Computation methods \\
\hline

FlashInfer, vLLM & \textbf{FlashInfer:} flashinfer.gemm, fused\_moe, quantization, topk\newline \textbf{vLLM:} KVCaching, parallelism & \textbf{FlashInfer:} Kernel Fusion, Conditional Computing, Low-bit Quantization methods\newline \textbf{vLLM:} Memory Management methods, Parallel Computation methods \\
\hline

FlashInfer, vLLM & \textbf{FlashInfer:} FlashInfer Attention Kernels, cascade, comm, norm, page, sampling, topk\newline \textbf{vLLM:} KVCaching, Multi-LoRA, Quantization, Speculative decoding, parallelism & \textbf{FlashInfer:} Tailored Attention, Cascade Inference, Parallel Computation methods, Kernel Fusion, Memory Management methods\newline \textbf{vLLM:} Memory Management methods, Network Pruning, Quantization, Speculative Decoding, Parallel Computation methods \\
\hline

FlashInfer, vLLM & \textbf{FlashInfer:} FlashInfer Attention Kernels, page, quantization\newline \textbf{vLLM:} Multi-LoRA, parallelism & \textbf{FlashInfer:} Tailored Attention, Memory Management methods, Low-bit Quantization methods\newline \textbf{vLLM:} Network Pruning, Parallel Computation methods \\
\hline

FlashInfer, vLLM & \textbf{FlashInfer:} norm\newline \textbf{vLLM:} Multi-LoRA & \textbf{FlashInfer:} Kernel Fusion\newline \textbf{vLLM:} Network Pruning \\
\hline

FlashInfer, vLLM & \textbf{FlashInfer:} FlashInfer Attention Kernels, green\_ctx\newline \textbf{vLLM:} Multi-LoRA & \textbf{FlashInfer:} Tailored Attention, Attention Simplification\newline \textbf{vLLM:} Network Pruning \\
\hline

FlashInfer, vLLM & \textbf{FlashInfer:} FlashInfer Attention Kernels, cascade, norm, page, sampling\newline \textbf{vLLM:} KVCaching, Multi-LoRA, parallelism & \textbf{FlashInfer:} Tailored Attention, Cascade Inference, Kernel Fusion, Memory Management methods\newline \textbf{vLLM:} Memory Management methods, Network Pruning, Parallel Computation methods \\
\hline

FlashInfer, vLLM & \textbf{FlashInfer:} FlashInfer Attention Kernels, cascade, comm, norm, page, sampling, topk\newline \textbf{vLLM:} CUDA, Chunked prefill, KVCaching, Multi-LoRA, Quantization, Speculative decoding, parallelism & \textbf{FlashInfer:} Tailored Attention, Cascade Inference, Parallel Computation methods, Kernel Fusion, Memory Management methods\newline \textbf{vLLM:} Kernel Optimizations methods, Request Scheduling methods, Memory Management methods, Network Pruning, Quantization, Speculative Decoding, Parallel Computation methods \\
\hline

FlashInfer, vLLM & \textbf{FlashInfer:} FlashInfer Attention Kernels, page, sampling\newline \textbf{vLLM:} CUDA, Chunked prefill, KVCaching, Multi-LoRA, Quantization, Speculative decoding, parallelism & \textbf{FlashInfer:} Tailored Attention, Memory Management methods\newline \textbf{vLLM:} Kernel Optimizations methods, Request Scheduling methods, Memory Management methods, Network Pruning, Quantization, Speculative Decoding, Parallel Computation methods \\
\hline

FlashInfer, vLLM & \textbf{FlashInfer:} FlashInfer Attention Kernels, page, quantization\newline \textbf{vLLM:} Multi-LoRA, Prefix caching, parallelism & \textbf{FlashInfer:} Tailored Attention, Memory Management methods, Low-bit Quantization methods\newline \textbf{vLLM:} Network Pruning, Memory Management methods, Parallel Computation methods \\
\hline

FlashInfer, vLLM & \textbf{FlashInfer:} page, sparse\newline \textbf{vLLM:} KVCaching, Prefix caching, parallelism & \textbf{FlashInfer:} Memory Management methods, Attention Simplification\newline \textbf{vLLM:} Memory Management methods, Parallel Computation methods \\
\hline

FlashInfer, vLLM & \textbf{FlashInfer:} FlashInfer Attention Kernels, page, sampling\newline \textbf{vLLM:} CUDA, Chunked prefill, Quantization, parallelism & \textbf{FlashInfer:} Tailored Attention, Memory Management methods\newline \textbf{vLLM:} Kernel Optimizations methods, Request Scheduling methods, Quantization, Parallel Computation methods \\
\hline

FlashInfer, vLLM & \textbf{FlashInfer:} Continuous batching, FlashInfer Attention Kernels, cascade, comm, flashinfer.gemm, fused\_moe, norm, page, quantization, sampling, topk\newline \textbf{vLLM:} CUDA, Chunked prefill, KVCaching, Multi-LoRA, Quantization, parallelism & \textbf{FlashInfer:} Tailored Attention, Cascade Inference, Parallel Computation methods, Kernel Fusion, Conditional Computing, Memory Management methods, Low-bit Quantization methods\newline \textbf{vLLM:} Kernel Optimizations methods, Request Scheduling methods, Memory Management methods, Network Pruning, Quantization, Parallel Computation methods \\
\hline

FlashInfer, vLLM & \textbf{FlashInfer:} FlashInfer Attention Kernels, page, sampling\newline \textbf{vLLM:} CUDA, Chunked prefill, KVCaching, Multi-LoRA, Prefix caching, Quantization, parallelism & \textbf{FlashInfer:} Tailored Attention, Memory Management methods\newline \textbf{vLLM:} Kernel Optimizations methods, Request Scheduling methods, Memory Management methods, Network Pruning, Quantization, Parallel Computation methods \\
\hline

SGLang, vLLM & \textbf{SGLang:} all parallelism, data parallelism, pipeline parallelism, tensor parallelism\newline \textbf{vLLM:} Multi-LoRA, parallelism & \textbf{SGLang:} Parallel Computation methods\newline \textbf{vLLM:} Network Pruning, Parallel Computation methods \\
\hline

SGLang, vLLM & \textbf{SGLang:} parallelism\newline \textbf{vLLM:} parallelism & \textbf{SGLang:} Parallel Computation methods\newline \textbf{vLLM:} Parallel Computation methods \\
\hline

SGLang, vLLM & \textbf{SGLang:} all parallelism, data parallelism, pipeline parallelism, tensor parallelism\newline \textbf{vLLM:} parallelism & \textbf{SGLang:} Parallel Computation methods\newline \textbf{vLLM:} Parallel Computation methods \\
\hline

SGLang, vLLM & \textbf{SGLang:} all parallelism, data parallelism, pipeline parallelism, tensor parallelism\newline \textbf{vLLM:} parallelism & \textbf{SGLang:} Parallel Computation methods\newline \textbf{vLLM:} Parallel Computation methods \\
\hline

SGLang, vLLM & \textbf{SGLang:} Continuous batching, parallelism\newline \textbf{vLLM:} parallelism & \textbf{SGLang:} Memory Management methods, Parallel Computation methods\newline \textbf{vLLM:} Parallel Computation methods \\
\hline

FlashInfer, SGLang & \textbf{FlashInfer:} Continuous batching, FlashInfer Attention Kernels, cascade, comm, flashinfer.gemm, fused\_moe, norm, page, quantization, topk\newline \textbf{SGLang:} Multi-LoRA, all parallelism, kv\_cache & \textbf{FlashInfer:} Tailored Attention, Cascade Inference, Parallel Computation methods, Kernel Fusion, Conditional Computing, Memory Management methods, Low-bit Quantization methods\newline \textbf{SGLang:} Request Scheduling methods, Parallel Computation methods, Memory Management methods \\
\hline

FlashInfer, SGLang & \textbf{FlashInfer:} FlashInfer Attention Kernels, cascade, flashinfer.gemm, fused\_moe, norm, page, quantization, topk\newline \textbf{SGLang:} Quantization, all parallelism, kv\_cache & \textbf{FlashInfer:} Tailored Attention, Cascade Inference, Kernel Fusion, Conditional Computing, Memory Management methods, Low-bit Quantization methods\newline \textbf{SGLang:} Low-bit Quantization methods, Parallel Computation methods, Memory Management methods \\
\hline

FlashInfer, SGLang & \textbf{FlashInfer:} FlashInfer Attention Kernels, cascade, flashinfer.gemm, fused\_moe, norm, page, quantization, topk\newline \textbf{SGLang:} Quantization, all parallelism, kv\_cache & \textbf{FlashInfer:} Tailored Attention, Cascade Inference, Kernel Fusion, Conditional Computing, Memory Management methods, Low-bit Quantization methods\newline \textbf{SGLang:} Low-bit Quantization methods, Parallel Computation methods, Memory Management methods \\
\hline

FlashInfer, SGLang & \textbf{FlashInfer:} FlashInfer Attention Kernels, cascade, flashinfer.gemm, fused\_moe, norm, page, quantization, topk\newline \textbf{SGLang:} Quantization, all parallelism, kv\_cache & \textbf{FlashInfer:} Tailored Attention, Cascade Inference, Kernel Fusion, Conditional Computing, Memory Management methods, Low-bit Quantization methods\newline \textbf{SGLang:} Low-bit Quantization methods, Parallel Computation methods, Memory Management methods \\
\hline

\end{longtable}
}
For the 18 repositories that combine FlashInfer and vLLM in Table \ref{tab:framework-combination-ratio} and \ref{tab:framework-method-category-combinations}, the main pattern is a cross-layer co-usage of FlashInfer’s kernel- and attention-oriented methods with vLLM’s serving methods. As shown in Table \ref{tab:framework-method-category-combinations}, FlashInfer APIs such as Attention Kernels, cascade, and fused-moe appear together with vLLM APIs or parameters such as KVCache, Chunked prefill, Prefix caching, Multi-LoRA, Quantization, Speculative decoding, and parallelism. The pattern of repositories that combine FlashInfer and vLLM suggests that these repositories use vLLM for serving-level execution, memory management, scheduling, and parallel inference, while relying on FlashInfer for lower-level attention and GPU-kernel optimization. This interpretation is consistent with the vLLM's design as a high-throughput distributed serving engine based on PagedAttention, block-level KV-cache memory management, and preemptive request scheduling \cite{vllm_document,kwon2023efficient} and FlashInfer's design as a customizable attention engine which is designed to integrate with serving frameworks such as vLLM \cite{kwon2023efficient,ye2025flashinfer}. 

\textbf{The highlighted methods in Table~\ref{tab:framework-combination-ratio} suggest that the co-usage of FlashInfer and vLLM is partly driven by complementary capabilities.} On the FlashInfer side, Tailored Attention appears in 15 of the 18 repositories and is frequently used with vLLM’s Parallel Computation methods in 14 repositories, Network Pruning in 13 repositories, and Memory Management methods and Quantization in 11 repositories each. Because each repository may use one or several serving methods at the same time, these counts are not mutually exclusive.
Cascade Inference appears in 6 repositories and is always used with vLLM’s Memory Management methods, Network Pruning, and Parallel Computation methods. Similarly, Kernel Fusion appears in 11 repositories and is mainly combined with vLLM’s Parallel Computation methods in 10 repositories, Memory Management methods and Network Pruning in 9 repositories each, and Quantization in 8 repositories. Conditional Computing appears in 4 repositories and is always used with vLLM’s Memory Management methods and Parallel Computation methods.

Repositories may have used one or more than one method at the same time .
These \textbf{results show that FlashInfer-specific methods are often paired with vLLM methods that support memory-aware, parallel, and optimized serving execution.} 
The highlighted vLLM methods including Request Scheduling, Network Pruning, and Speculative Decoding, also show a clear pairing with FlashInfer’s attention- and kernel-level methods. Request Scheduling appears in 7 repositories and is always used with FlashInfer’s Tailored Attention and Memory Management methods, while also appearing with FlashInfer’s Cascade Inference, Parallel Computation methods, and Kernel Fusion in 4 repositories each. Network Pruning appears in 14 repositories and is mostly combined with FlashInfer’s Tailored Attention in 13 repositories, Memory Management methods in 12 repositories, and Kernel Fusion in 9 repositories. Speculative Decoding appears in 6 repositories and is always used with FlashInfer’s Tailored Attention, while also appearing with FlashInfer’s Kernel Fusion and Memory Management methods in 5 repositories each. Finally, vLLM’s Parallel Computation methods appear in 16 repositories and are frequently used with FlashInfer’s Tailored Attention and Memory Management methods in 14 repositories each and Kernel Fusion in 10 repositories.

\textbf{Overall, this pattern suggests that repositories combining FlashInfer and vLLM may use vLLM for serving-level orchestration, such as scheduling, memory management, pruning, speculative decoding, and parallel execution, while using FlashInfer for specialized attention, cascade inference, kernel fusion, and conditional-computation primitives.} 
This interpretation could be due to vLLM’s design as a high-throughput serving engine based on PagedAttention, KV-cache memory management, and request scheduling, and with FlashInfer’s role as an attention and GPU-kernel optimization engine for LLM inference serving \cite{kwon2023efficient,ye2025flashinfer,vllm_document,FlashInfer_document}.

Based on Tables \ref{tab:framework-combination-ratio} and \ref{tab:framework-method-category-combinations}, \textbf{across the five repositories that combine SGLang and vLLM, the main recurring pattern is the joint use of Parallel Computation methods: all five repositories use vLLM’s `parallelism`, and all five also use SGLang parallelism-related APIs or parameters.} This pattern may suggest that the SGLang--vLLM combination is mainly used in repositories focused on scalable and distributed serving execution, where SGLang provides explicit parallelism controls and vLLM contributes high-throughput serving capabilities based on efficient memory management, scheduling, and batching \cite{zheng2024sglang,kwon2023efficient}. Table \ref{tab:repo-framework-methods-cleaned} also shows two additional patterns: one repository combines SGLang’s Continuous batching with vLLM’s parallelism, and one repository combines vLLM’s Network Pruning, with SGLang’s distributed parallelism methods. Overall, \textbf{these combinations may unlock more efficient serving for workloads that require both distributed parallel execution and support for multiple specialized versions of the same model, rather than only scaling a single base model \cite{zheng2024sglang,kwon2023efficient,vllm_document,sglang_document}. }

Across the four repositories that combine FlashInfer and SGLang, FlashInfer’s Tailored Attention, Cascade Inference, Kernel Fusion, and Conditional Computing methods appear in all four repositories. In each of these repositories, these FlashInfer-specific methods are used together with SGLang’s Memory Management methods, mainly kv-cache, and SGLang’s Parallel Computation methods. In three of the four repositories, these FlashInfer-specific methods are also used with SGLang’s Low-bit Quantization methods. This pattern suggests that the FlashInfer--SGLang combination is mostly associated with using FlashInfer’s attention- and kernel-level optimization methods alongside SGLang’s runtime-level capabilities for cache management, parallel execution, and, in most cases, quantized serving.


One important finding of this section is that the \textbf{studied LLM serving frameworks are not frequently used together within the same repositories.} This suggests that, in practice, developers often rely on a single serving framework rather than combining multiple frameworks and their methods. However, the reasons behind this limited co-usage remain unclear and should be investigated in future studies. Possible explanations that need to be investigated in the future research include the complexity of integrating different serving frameworks, compatibility constraints between specific frameworks and LLMs, and the lack of clear documentation or examples showing how different frameworks and their serving methods can be combined in practice. Therefore, future work should examine the technical and practical barriers that may prevent developers from adopting multi-framework serving configurations when separate tools provide different complementary efficient serving methods.
\textbf{The findings of Section \ref{rq2.2} show that when frameworks are combined, they are often used to connect complementary capabilities across different layers of the serving stack.} For example, the FlashInfer and vLLM combination links vLLM’s serving-level methods, such as memory management, scheduling, and parallel execution, with FlashInfer’s attention- and kernel-level optimizations. \textbf{These findings can help practitioners better understand which framework combinations are used in practice and can guide framework maintainers to provide clearer integration support, examples, and documentation. }They also motivate future survey research on why multi-framework adoption remains uncommon.

\subsubsection{What serving frameworks are adopted by developers for each type of LLMs?}
\label{rq2.3}
In this section, we analyze the LLMs associated with each serving framework along four model-level dimensions: model family, model parameter-size category, model specialization, and model deployment context. These dimensions help characterize which types of LLMs are observed with each framework in the studied GitHub repositories, including the model family they belong to, their reported parameter scale, the task or domain for which they are specialized, and the practical context in which they are deployed. Table \ref{tab:model_categories} summarizes these categories, together with their descriptions and subcategories. Furthermore, Table \ref{tab:combinationsofmodelsandframeworkds_restructured} presents detailed information on the categories and subcategories of LLMs used with each studied framework. 

\begin{table}[!htbp]
\centering
\caption{LLM model categories, subcategories, and their descriptions}
\label{tab:model_categories}

\small
\renewcommand{\arraystretch}{1.15}
\setlength{\tabcolsep}{3pt}

\begin{tabularx}{\textwidth}{
    |>{\raggedright\arraybackslash}p{2.2cm}
    |>{\raggedright\arraybackslash}p{3.4cm}
    |>{\raggedright\arraybackslash}X|
}
\hline
\textbf{Category} & \textbf{Description} & \textbf{Subcategories} \\ \hline

Model Family & Identifies the model family or provider lineage associated with the LLM & Baichuan, InternVL, LLaMA, Qwen, ChatGLM, DeepSeek, InternLM, LLaVA, Vicuna \\ \hline


Size & Captures the reported parameter scale of the LLM &Small models upto 2B, Medium models 3B to 9B, Large models 10B to 34B, Very large models 70B plus, Mixture of Experts (MoE) models \\ \hline

Specialization & Describes the main task, domain, or capability for which the LLM is designed or commonly used & General chat assistants, Instruction following, Vision language models, Code generation, Mathematics reasoning, Embedding and retrieval, Grounding and long context, General base models\\ \hline

Deployment Context & Indicates the practical setting in which the LLM is used &Production ready chat models, Instruction serving and agents, Multimodal application models, Research and base models, Edge and resource constrained, Domain specific pipelines \\ \hline

\end{tabularx}

\end{table}

\newcommand{\us}{\_\hspace{0pt}}

\begin{table*}[!htbp]
\centering
\caption{Detailed information on the LLMs used with the LLM serving frameworks.}
\label{tab:combinationsofmodelsandframeworkds_restructured}
\scriptsize
\setlength{\tabcolsep}{2pt}
\renewcommand{\arraystretch}{1.05}

\begin{tabularx}{\textwidth}{|
>{\RaggedRight\arraybackslash}p{1.85cm}|
>{\RaggedRight\arraybackslash}X|
>{\RaggedRight\arraybackslash}X|
>{\RaggedRight\arraybackslash}X|
>{\RaggedRight\arraybackslash}X|
>{\RaggedRight\arraybackslash}p{1.5cm}|}
\hline
\textbf{Category} & \textbf{vLLM} & \textbf{SGLang} & \textbf{TensorRT-LLM} & \textbf{LMDeploy} & \textbf{FlashInfer} \\
\hline

\textbf{Model Family} &
LLaMA family \newline
Qwen family \newline
GLM family \newline
Vicuna family \newline
InternLM family &
LLaMA family \newline
Qwen family &
LLaMA family \newline
LLaVA family \newline
Vicuna family &
Qwen family \newline
InternLM family \newline
LLaMA family \newline
Baichuan family \newline
InternVL family \newline
ChatGLM family \newline
Vicuna family \newline
DeepSeek family &
LLaMA family \\
\hline

\textbf{Size} &
Medium models 3B to 30B \newline
Small models upto 3B \newline
Large models 30B to 100B \newline
Very large models 70B plus &
Medium models 3B to 9B \newline
Small models upto 2B \newline
MoE models &
Medium models 3B to 9B &
Medium models 3B to 9B \newline
Large models 10B to 34B \newline
Small models upto 2B \newline
Very large models 70B plus \newline
MoE models &
MoE models \\
\hline

\textbf{Specialization} &
Instruction following \newline
General chat assistants \newline
Vision language models \newline
General base models \newline
Code generation &
Instruction following \newline
Vision language models \newline
General chat assistants &
General chat assistants \newline
Instruction following \newline
Vision language models &
General chat assistants \newline
Vision language models \newline
General base models \newline
Instruction following \newline
Embedding and retrieval \newline
Mathematics reasoning \newline
Grounding and long context \newline
Code generation &
Instruction following \\
\hline

\textbf{Deployment Context} &
Instruction serving and agents \newline
Production ready chat models \newline
Multimodal application models \newline
Research and base models \newline
Domain specific pipelines &
Instruction serving and agents \newline
Multimodal application models \newline
Production ready chat models &
Production ready chat models \newline
Instruction serving and agents \newline
Multimodal application models &
Production ready chat models \newline
Multimodal application models \newline
Edge and resource constrained \newline
Research and base models \newline
Instruction serving and agents \newline
Domain specific pipelines &
Instruction serving and agents \\
\hline

\end{tabularx}

\end{table*}

Identifying the \textbf{Model Family} helps determine whether a serving framework is adopted across a diverse set of LLM ecosystems or is mainly associated with a narrower group of model families.
\textbf{As the \textbf{Model Family} category shows, vLLM and LMDeploy are used with a broader range of model families, whereas FlashInfer is used only with the LLaMA family. This result suggests that vLLM and LMDeploy currently demonstrate broader popularity and compatibility across diverse LLM families.} Some LLMs are used exclusively with specific frameworks. For instance, Baichuan, ChatGLM, DeepSeek, and InternVL are used only with LMDeploy, while LLaVA is used only with TensorRT-LLM. 
In contrast, several LLMs are adopted across multiple frameworks. Notably, LLaMA is used with all frameworks, Qwen is used with vLLM, SGLang, and LMDeploy, InternLM is used with vLLM and LMDeploy, and Vicuna is used with vLLM, TensorRT-LLM, and LMDeploy.

The \textbf{size} category is useful because model size is related to compute demand and deployment difficulty \cite{sheng2023flexgen,gao2024displlm}. Knowing whether frameworks are used with small, medium, large, very large, or MoE models can help explain which frameworks are adopted for lightweight deployment versus larger and more resource-demanding LLM serving. 
The description of each subcategory in the category \textbf{size} is provided in what follows. \textit{Small models up to 2B} have up to about 2B parameters and are typically designed for lightweight or resource-efficient use. \textit{Medium models 3B to 9B} contain about 3B to 9B parameters and usually balance capability with deployment efficiency. \textit{Large models 10B to 34B} contain about 10B to 34B parameters and are generally aimed at stronger performance on complex tasks. \textit{Very large models 70B plus} have 70B or more parameters and are typically used when maximum capability is prioritized over cost. Finally, \textit{MoE models} use a mixture-of-experts architecture, where only part of the model is activated for each input to improve scaling efficiency.
As shown in the \textbf{size} category of Table \ref{tab:combinationsofmodelsandframeworkds_restructured}, large and very large LLMs are used only with vLLM and LMDeploy. This \textbf{finding suggests that developers aiming to serve large or very large LLMs may consider vLLM and LMDeploy as two strong candidate frameworks when productionizing their systems.}

Category \textbf{Specialization} helps connect each model to the main task, domain, or capability it is intended or commonly documented to support, such as dialogue, instruction following, vision-language understanding, code-related tasks, mathematical reasoning, embedding and retrieval, grounding with external or long context, or downstream adaptation. \textit{General chat assistants} refer to models designed for dialogue or assistant-style use case. \textit{Instruction following} models refer to models trained or aligned to follow user intent and task-oriented prompts. \textit{Vision language models} refer to models that connect visual and language inputs for multimodal understanding. \textit{Code generation} models refer to models designed for code-related tasks such as program synthesis, code completion, debugging, and code infilling. \textit{Mathematics reasoning} models refer to models designed or evaluated for mathematical and quantitative problem solving with step-by-step solutions. \textit{Embedding and retrieval} models refer to models that encode text into vector representations for tasks such as retrieval, reranking, clustering, and classification. \textit{Grounding and long context} models refer to models or systems that use retrieved external knowledge or extended input contexts to support response generation. Finally, \textit{General base models} refer to broadly trained foundation models that can be adapted to downstream tasks through fine-tuning or other adaptation methods.

Category \textbf{specialization}  of Table \ref{tab:combinationsofmodelsandframeworkds_restructured} shows that \textbf{LMDeploy and vLLM are used with a broader variety of LLMs specialized for different domains, including mathematical reasoning, code generation, and vision-language tasks. In contrast, FlashInfer is used only with LLMs specialized for instruction following.}

\cite{paleyes2022challenges,jouini2024survey}
.

This helps interpret framework usage from a software-engineering perspective, rather than only from the perspective of serving-method usage. The description of each subcategory in the category Deployment Context is provided in what follows. \textit{Production ready chat models} are models typically used in deployed chat assistants and user-facing conversational applications. \textit{Instruction serving and agents} are models typically used in systems that execute explicit instructions or power agent-style workflows. \textit{Multimodal application models} are models typically used in applications that process both text and visual inputs, such as images, documents, or charts. \textit{Research and base models} are models typically used as base or experimental models for further fine-tuning, evaluation, or research. \textit{Edge and resource constrained} models are typically used in low-memory or limited-compute environments where efficient deployment is important. Finally, \textit{Domain specific pipelines} are models typically used in specialized pipelines targeting particular tasks such as coding, math, retrieval, or grounded generation.
As shown in the \textbf{deployment context} category of Table \ref{tab:combinationsofmodelsandframeworkds_restructured}, LMDeploy is the only framework used with edge and resource-constrained LLMs. This suggests that \textbf{developers aiming to serve LLMs designed for low-memory or limited-compute environments, where deployment efficiency is particularly important, may consider LMDeploy a strong candidate. In addition, only vLLM and LMDeploy are used with research and base models. Therefore, researchers seeking serving frameworks for such models may consider these two tools as strong candidate choices.}

The findings of Table~\ref{tab:combinationsofmodelsandframeworkds_restructured} show that only LMDeploy is associated with highly specialized use cases, such as embedding retrieval, math reasoning, and specialized grounding. 
Furthermore, \textbf{vLLM appears to balance breadth and practicality, as it is used with several model families, multiple specializations, several sizes, and multiple deployment contexts.} In contrast, SGLang appears to be more concentrated around mainstream instruction/chat workloads and small- to medium-sized LLM adoption.
The findings of Table~\ref{tab:combinationsofmodelsandframeworkds_restructured} can benefit several stakeholder groups. \textbf{Developers of LLM-based applications can use these results to identify which serving frameworks are used with specific model} families, model sizes, specializations, and deployment contexts, which may help narrow the set of frameworks to evaluate. 
Developers and \textbf{maintainers of LLM serving frameworks can use the findings to understand where their frameworks show broad or limited observed coverage and to improve support}, documentation, and examples for underrepresented model categories.
MLOps, platform, and infrastructure engineers can also use the results to plan deployment environments based on the types of models served with each framework, especially for large models, multimodal models, MoE models, and edge or resource-constrained settings. 
Finally, researchers in software engineering and AI systems can use these findings to study how model characteristics are associated with framework adoption in open-source LLM-serving projects.


\begin{tcolorbox}[breakable]
\textbf{Finding 2.} Developers combine multiple serving methods within the same framework to address different serving bottlenecks together, whereas multi-framework usage remains limited. vLLM has the largest absolute co-usage count, while FlashInfer’s highest co-usage ratio indicates that it is more often adopted in multi-framework settings. Recurring framework combinations connect complementary capabilities across the serving stack, including scheduling, memory management, parallel execution, and attention- or kernel-level optimization. Framework adoption also varies across LLM types, with vLLM and LMDeploy used across broader model-family ranges than other frameworks.
\end{tcolorbox}

\subsection{RQ3. How are LLM serving frameworks used across different types of LLM-based software systems?}
This section presents the characteristics and topics of repositories that use serving frameworks.

\subsubsection{What types of LLM-based software systems adopt LLM serving frameworks in terms of repository intent, technical focus, use case, and system design?}
\label{rq3.1}
This section presents the results of clustering and topic modeling conducted to identify the main topics of repositories that use LLM serving frameworks from four perspectives: Repository Intent, Technical Focus, Primary Use Cases, and System Design, as described in Section~\ref{Summ_clustering}.
Tables \ref{tab:repository_intent_topics}, \ref{tab:technical_focus_topics}, \ref{tab:primary_use_cases_topics}, and \ref{tab:system_design_topics} presents the topics identified for Repository Intent, Technical Focus, Primary Use Cases, and System Design, respectively, together with a description of each topic in the \textit{Description} column. 
As Tables \ref{tab:repository_intent_topics}, \ref{tab:technical_focus_topics}, \ref{tab:primary_use_cases_topics}, and \ref{tab:system_design_topics} show, the numbers of identified topics for Repository Intent, Technical Focus, Primary Use Cases, and System Design are 10, 7, 11, and 9, respectively.

Table \ref{tab:repository_intent_topics} shows the intent of repositories that use the studied serving frameworks. \textbf{The results presented in Table \ref{tab:repository_intent_topics} reveal that repositories using LLM serving frameworks span a diverse set of intents. The discovered repository-intent topics cover high-performance serving infrastructure, RL workflows, retrieval and RAG systems, multimodal understanding, chatbot and agent systems, speech applications, and visual grounding tasks. 
These findings suggest that LLM serving frameworks are used in general-purpose components for building and deploying complex AI systems.}

\begin{table}[!htbp]
\centering
\caption{Topic modeling results for the \textit{Repository Intent} summary category.}
\label{tab:repository_intent_topics}
\footnotesize
\renewcommand{\arraystretch}{1.2}
\setlength{\tabcolsep}{4pt}
\begin{tabularx}{\textwidth}{|p{0.7cm}|p{3.2cm}|X|}
\hline
\textbf{Topic No.} & \textbf{Topic Name} & \textbf{Description} \\
\hline
0 & High-Performance GPU Serving & Repos focused on high-performance GPU-based serving, optimization, and distributed deployment of DL systems. \\
\hline
1 & RL Frameworks & Repos providing flexible tools, libraries, and agent-based frameworks for RL and post-training workflows. \\
\hline
2 & Mathematical Reasoning with RL & Repos centered on RL for mathematical reasoning, including dataset preparation and self-training or verification processes. \\
\hline
3 & Retrieval and RAG Systems & Repos aimed at information retrieval, question answering, and benchmarking using evaluation metrics and RAG-based approaches. \\
\hline
4 & Speech and Voice Applications & Repos developing speech and voice systems, including recognition, text-to-speech, and real-time multimodal interaction. \\
\hline
5 & General ML and Generative AI Projects & Repos showcasing or implementing machine learning and DL projects, including optimization and generative AI applications. \\
\hline
6 & Multimodal Vision Understanding & Repos focused on multimodal video and image understanding, including benchmarking, tuning, and generation tasks. \\
\hline
7 & Chatbot and Agent Systems & Repos designed for chatbot systems, enabling interaction, agent functionality, personalization, and automation outputs. \\
\hline
8 & Data Generation and Training Pipelines & Repos emphasizing code, datasets, and pipelines for data generation, preprocessing, and reproducible training workflows. \\
\hline
9 & Visual Grounding and Reasoning & Repos targeting visual grounding and reasoning in vision-language systems, including image-based understanding and learning. \\
\hline
\end{tabularx}

\end{table}

Topic \textit{High-Performance GPU Serving} in Table \ref{tab:repository_intent_topics} highlights the importance of efficient serving, distributed deployment, and GPU optimization in LLM-based software systems. 
The existence of a distinct topic around GPU serving suggests that infrastructure optimization is important for developers adopting LLM serving frameworks. 
Another observation of Table \ref{tab:repository_intent_topics} is the presence of reinforcement-learning-related repositories. The identified topics include both general \textit{RL frameworks} and repositories specifically targeting \textit{mathematical reasoning with RL}. 
\textbf{These findings indicate that LLM serving frameworks are integrated into post-training\cite{van2025post} and reasoning-oriented workflows rather than being used only for inference.} In particular, the existence of a dedicated topic for \textit{Mathematical Reasoning with RL} suggests interest in using RL to improve reasoning capabilities and self-verification processes in LLM systems\cite{wang2025reinforcement}. The results of Table \ref{tab:repository_intent_topics} also show that retrieval-oriented systems represent a category of repositories using LLM serving frameworks. The \textit{Retrieval and RAG Systems} topic includes repositories focused on information retrieval, question answering, benchmarking, and retrieval-augmented generation approaches. This finding suggests that integrating external knowledge sources and retrieval mechanisms has occured in LLM-based systems.

Furthermore, several identified topics in Table \ref{tab:repository_intent_topics} involve multimodal understanding, speech interaction, visual grounding, and vision-language reasoning. In particular, the topics \textit{Speech and Voice Applications}, \textit{Multimodal Vision Understanding}, and \textit{Visual Grounding and Reasoning} indicate that repositories support systems operating across multiple modalities, including text, image, audio, and video. 
These findings suggest that LLM-based systems are evolving beyond text-only interaction toward more general multimodal AI systems. Another interesting finding of Table \ref{tab:repository_intent_topics} is the presence of a dedicated topic for \textit{Chatbot and Agent Systems}. 
Repositories within this topic focus on interaction, personalization, automation, and agent functionality. 
This \textbf{result indicates that agent-oriented applications form a category of systems using LLM serving frameworks.} 
It also suggests that developers combine LLM serving capabilities with autonomous functionalities. 
The identified topic \textit{Data Generation and Training Pipelines} of Table \ref{tab:repository_intent_topics} further suggests that repositories using LLM serving frameworks integrate end-to-end engineering workflows. These repositories emphasize dataset preparation, preprocessing pipelines, reproducible training workflows, and data generation processes. This observation indicates that LLM serving frameworks are used as components within broader AI development pipelines rather than as isolated serving modules.

In summary, \textbf{the identified topics listed in Table \ref{tab:repository_intent_topics} demonstrate that repositories using LLM serving frameworks combine both infrastructure oriented and application oriented concerns.} Some topics focus primarily on deployment efficiency, distributed serving, and GPU optimization, while others focus on application-level functionalities such as retrieval systems, multimodal interaction, chatbot agents, and reasoning systems. This suggests that LLM-based software systems that use the studied serving frameworks span multiple abstraction layers, ranging from low-level serving infrastructure to high-level intelligent applications. Overall, \textbf{the results of Table \ref{tab:repository_intent_topics} suggest that the ecosystem of repositories using LLM serving frameworks is highly heterogeneous and multidisciplinary. }

The findings of Table \ref{tab:repository_intent_topics} can benefit multiple stakeholders, including serving-framework developers, researchers, and technology decision-makers. \textbf{Developers of LLM serving frameworks can use these findings to better understand the practical contexts in which their frameworks are used, and to improve support for those contexts.} \textbf{Besides, framework builders such as orchestration-tool developers, and middleware providers can use these findings to improve compatibility and integration across the broader LLM stack, especially because the results suggest that serving frameworks are used as part of larger end-to-end AI pipelines rather than as isolated components.} 
\textbf{Researchers in software engineering and AI systems can use these findings to understand how LLM serving frameworks are being adopted in practice}, to identify emerging application domains, and to motivate future research on system support for multimodal, retrieval-enhanced, agent-oriented, and reasoning-intensive workflows. Technology decision-makers can use these findings to better understand the practical landscape of LLM-based software systems, which can help them prioritize investments, evaluate technical directions, and make more informed decisions about infrastructure.

Table \ref{tab:technical_focus_topics} shows the technical focus of repositories that use the studied serving frameworks. The results in Table \ref{tab:technical_focus_topics} show that the technical focus of GitHub repositories using LLM serving frameworks is broad. The identified topics span both model-centric concerns, such as training, tuning, and reinforcement-learning-related workflows, and systems-centric concerns, such as CUDA-based optimization, API deployment, containerization, and Kubernetes based infrastructure. This indicates that \textbf{repositories using LLM serving frameworks do not focus only on inference execution, but rather operate across the broader technical stack of LLM-based software systems.}

\begin{table}[!htbp]
\centering
\caption{Topic modeling results for the \textit{Technical Focus} summary category.}
\label{tab:technical_focus_topics}
\footnotesize
\renewcommand{\arraystretch}{1.2}
\setlength{\tabcolsep}{4pt}
\begin{tabularx}{\textwidth}{|p{0.7cm}|p{3.2cm}|X|}
\hline
\textbf{Topic No.} & \textbf{Topic Name} & \textbf{Description} \\
\hline
0 & Model Training and Optimization & Repos focused on training, evaluation, optimization, tuning, and performance measurement for learning and generation systems. \\
\hline
1 & CUDA Parallelism and Kernel Optimization & Repos centered on CUDA-based parallelism techniques, including tensor parallelism, pipeline parallelism, batching, kernel optimization, memory management, and attention computation. \\
\hline
2 & Containerized API Deployment & Repos focused on Python-based API and containerized deployment workflows, including Docker, PyTorch, FastAPI, OpenAI integration, image handling, and CUDA support. \\
\hline
3 & Python Environment and Serving Workflows & Repos concerned with Python environments, dependencies, installation, build processes, scripting, evaluation, and serving workflows. \\
\hline
4 & Speech, OCR, and Multimedia Processing Pipelines & Repos focused on audio, speech, and voice processing, along with object detection, OCR-related components, and transport or connector-based processing pipelines. \\
\hline
5 & Hugging Face and RL Integration & Repos centered on Hugging Face integration, device mapping, resharding, and RL-related actor or GRPO workflows. \\
\hline
6 & Kubernetes-Based AI Deployment & Repos focused on Kubernetes-based AI deployment and configuration, including provisioning, generation services, metric collection, FAISS integration, and environment setup. \\
\hline
\end{tabularx}

\end{table}

One finding of Table \ref{tab:technical_focus_topics} is that the topics \textit{Model Training and Optimization} and \textit{CUDA Parallelism and Kernel Optimization} show that repositories emphasize training efficiency, batching, tensor parallelism, pipeline parallelism, kernel optimization, memory management, and attention computation. 
This suggests that performance and optimization are technical concerns in repositories using LLM serving frameworks. Another important finding of Table \ref{tab:technical_focus_topics} is that deployment engineering is a distinct technical focus. The topics \textit{Containerized API Deployment} and \textit{Kubernetes-Based AI Deployment} indicate that repositories include practical deployment workflows based on Docker, FastAPI. This suggests that LLM serving frameworks are also used in repositories targeting deployable and operational AI systems. 
The identified topic \textit{Python Environment and Serving Workflows} in Table \ref{tab:technical_focus_topics} further shows that software environment management is itself a technical concern. 
The emphasis on dependencies, installation, build processes, scripting, evaluation, and serving workflows suggests that repositories using LLM serving frameworks require engineering effort for execution management. Besides, Table \ref{tab:technical_focus_topics} shows that the technical focus of these repositories extends beyond text-only pipelines. The topic \textit{Speech, OCR, and Multimedia Processing Pipelines} indicates that repositories integrate audio, speech, OCR, and other multimedia components into their processing workflows. This suggests that LLM serving frameworks are used within multimodal AI systems. 
Another result of Table \ref{tab:technical_focus_topics} is the presence of a topic on \textit{Hugging Face and RL Integration} which suggests that LLM serving frameworks are not used in isolation, but are sometimes embedded within broader post-training pipelines\cite{van2025post}.

Taken together, the identified topics listed in Table \ref{tab:technical_focus_topics} suggest that \textbf{the technical focus of repositories using LLM serving frameworks combines optimization, deployment, environment management, ecosystem integration, and multimodal processing.} In other words, these repositories reflect not only low-level serving concerns, but also the end-to-end technical requirements of building, deploying, and maintaining practical LLM-based systems. 
A further implication of the Table \ref{tab:technical_focus_topics} findings is that repositories using LLM serving frameworks span multiple layers of abstraction. Some topics focus on low-level computational efficiency, such as CUDA kernels, parallelism, and memory management, whereas others focus on higher-level engineering concerns, such as containerization. This suggests that \textbf{modern LLM serving frameworks function as part of a larger technical ecosystem rather than as standalone inference utilities.}


\textbf{The findings of Table \ref{tab:technical_focus_topics} can benefit the developers of LLM-based applications choose frameworks that better match their technical needs. 
Also, it can benefit the developers of LLM serving frameworks prioritize features and integrations that better reflect practical usage contexts.} 
Since the identified topics include CUDA-based optimization, containerized deployment, Kubernetes-based deployment, Hugging Face integration, RL-related workflows, and multimedia pipelines, the results can guide framework developers in improving support for various identified end-to-end engineering needs rather than focusing only on core inference performance. 
For MLOps engineers, these findings can help them improve deployment pipelines and operational support for LLM systems. 
The presence of  topics related to containerized APIs, Kubernetes-based deployment, dependency management, and execution workflows suggests that LLM systems require robust deployment, orchestration, and reproducibility support. These results can therefore help such stakeholders improve system maintainability, and deployment efficiency. 
For framework builders and platform providers, the findings can help them improve interoperability across the broader LLM engineering stack. 
The identified topic on Hugging Face and RL integration, together with topics on deployment workflows and multimedia pipelines, suggests that serving frameworks are used together with other model-development and deployment ecosystems. 
This can help framework providers identify where better compatibility, integration support, and modular interfaces may be most useful. 
For researchers, the findings can help them better understand the practical technical landscape of repositories using LLM serving frameworks. 
In particular, \textbf{the results provide empirical evidence that modern repositories combine optimization, deployment, environment management, post-training integration, and multimodal processing.} This can help researchers identify underexplored technical challenges and motivate future work on end-to-end LLM software engineering across different applications.

Table \ref{tab:primary_use_cases_topics} shows the primary use cases of repositories that use the studied serving frameworks. The topics of Table \ref{tab:primary_use_cases_topics} indicate that repositories using LLM serving frameworks support several practical AI use cases rather than a single narrow use case.
\begin{table}[!htbp]
\centering
\caption{Topic modeling results for the \textit{Primary Use Cases} summary category.}
\label{tab:primary_use_cases_topics}
\footnotesize
\renewcommand{\arraystretch}{1.2}
\setlength{\tabcolsep}{4pt}
\begin{tabularx}{\textwidth}{|p{0.7cm}|p{3.2cm}|X|}
\hline
\textbf{Topic No.} & \textbf{Topic Name} & \textbf{Description} \\
\hline
0 & Multimodal Generation and Understanding & Repos focused on multimodal image, text, and video generation, understanding, and real-time recognition tasks. \\
\hline
1 & Efficient LLM Serving & Repos designed for running and serving LLMs efficiently, including multimodal LLM applications. \\
\hline
2 & Machine Learning Training and Usage & Repos supporting machine learning models and datasets for learning, training, and model usage. \\
\hline
3 & Cloud-Based AI Deployment & Repos focused on AI model training, inference, and deployment across cloud-based systems and environments. \\
\hline
4 & Long-Context QA and Benchmarking & Repos centered on language models for long-context understanding, question answering, and benchmarking tasks. \\
\hline
5 & Mathematical Reasoning Systems & Repos targeting mathematical reasoning, problem solving, and evaluation of reasoning tasks. \\
\hline
6 & Model Benchmarking and Fine-Tuning & Repos focused on model benchmarking, tuning, fine-tuning, evaluation, and inference across systems. \\
\hline
7 & RL Workflows & Repos implementing RL algorithms, including training and fine-tuning workflows. \\
\hline
8 & Agent-Based Systems & Repos designed for agent-based systems, including tools, tool-calling, and handling complex tasks. \\
\hline
9 & Retrieval and Recommendation Systems & Repos focused on information retrieval, document search, text extraction, and recommendation systems. \\
\hline
10 & GPU-Optimized Model Deployment & Repos focused on GPU-based model deployment and optimized execution, including kernel-level and single-node serving systems. \\
\hline
\end{tabularx}

\end{table}
One finding of Table \ref{tab:primary_use_cases_topics} is that \textit{efficient LLM serving} itself forms a use-case category. This suggests that efficient serving of LLMs constitute an application goal in their own. The presence of a separate topic for \textit{GPU-optimized model deployment} further reinforces that optimized execution, is an important practical use case for repositories adopting LLM serving frameworks. Another result of Table \ref{tab:primary_use_cases_topics} is the presence of multimodal use cases. The topic \textit{Multimodal Generation and Understanding} covers image, text, and video generation, understanding, and real-time recognition tasks. This indicates that \textbf{repositories using LLM serving frameworks are not limited to text-only systems, but are used to support multimodal applications that integrate multiple input and output modalities.} In the Table \ref{tab:primary_use_cases_topics}, the topics \textit{Machine Learning Training and Usage} and \textit{Model Benchmarking and Fine-Tuning} indicate that there are repositories supporting training, benchmarking, evaluation, and fine-tuning activities. This suggests that \textbf{LLM serving frameworks are also used within end-to-end AI workflows that connect model development, evaluation, and operational use, rather than being limited to final-stage inference alone.} 

Furthermore, the identified topics of Table \ref{tab:primary_use_cases_topics} also highlight the importance of knowledge-intensive and retrieval-oriented applications. The topics \textit{Long-Context QA and Benchmarking} and \textit{Retrieval and Recommendation Systems} show that repositories support question answering, document search, text extraction, benchmarking, and recommendation tasks. This suggests that some repositories using LLM serving frameworks build systems that rely on long-context processing, retrieval mechanisms, and information-access workflows. Another result of Table \ref{tab:primary_use_cases_topics} is the presence of  use-case categories for mathematical reasoning and RL workflows. \textbf{The topics \textit{Mathematical Reasoning Systems} and \textit{RL Workflows} indicate there are repositories that use LLM serving frameworks for reasoning-oriented tasks and RL-based training or post-training \cite{van2025post}.} 
The topic Agent-Based Systems in Table \ref{tab:primary_use_cases_topics} provides evidence that the studied frameworks are also used in interactive systems. Since this topic includes tools, tool-calling, and handling complex tasks, it indicates that there are repositories that employ LLM serving frameworks as part of agent-oriented systems.

\textbf{For developers of LLM serving frameworks can benefit from the results of the Table \ref{tab:primary_use_cases_topics} to prioritize improvements for the use cases that appear in practice and broaden framework support beyond standalone inference.} 
Also, the results can guide framework developers toward improving support for diverse downstream workloads and end-to-end usage settings. For researchers, the findings of Table \ref{tab:primary_use_cases_topics} can help them understand the practical use-case landscape of repositories using LLM serving frameworks and identify important directions for future empirical and technical research.

Table \ref{tab:system_design_topics} shows the system design of repositories that use the studied serving frameworks. 

\begin{table}[!htbp]
\centering
\caption{Topic modeling results for the \textit{System Design} summary category.}
\label{tab:system_design_topics}
\footnotesize
\renewcommand{\arraystretch}{1.2}
\setlength{\tabcolsep}{4pt}
\begin{tabularx}{\textwidth}{|p{0.7cm}|p{3.2cm}|X|}
\hline
\textbf{Topic No.} & \textbf{Topic Name} & \textbf{Description} \\
\hline
0 & Scalable Serving Backends & Repos focused on scalable backend system design for serving and execution, including API integration, resource management, and performance optimization strategies. \\
\hline
1 & Containerized GPU Deployment & Repos designed around containerized and GPU-enabled deployment using Docker, cloud services, multi-GPU setups, and service interfaces. \\
\hline
2 & Python/Conda Pipeline Organization & Repos structured around Python and Conda environments with organized directories, scripts, and pipelines for training, dataset handling, and evaluation. \\
\hline
3 & Dependency and Script Management & Repos emphasizing dependency management, documentation, community contributions, and command-based script execution workflows. \\
\hline
4 & Microservices and Cloud Infrastructure & Repos implementing microservices architectures with containerization, orchestration, load balancing, fault tolerance, and scalable cloud infrastructure. \\
\hline
5 & Experiment Tracking and GPU Resource Management & Repos focused on experiment tracking, GPU resource utilization, scalable mapping, and seamless integration policies for managing experiments. \\
\hline
6 & CUDA and Tensor Parallelism Frameworks & Repos centered on tensor operations, CUDA kernels, parallelism, and efficient memory management within frameworks like PyTorch. \\
\hline
7 & Modular Training Pipelines & Repos organized as modular training pipelines with script execution, evaluation integration, and codebase compatibility handling. \\
\hline
8 & Kubernetes-Based Cloud Infrastructure & Repos designed for Kubernetes-based cloud infrastructure, including cluster scaling, job management, Helm usage, and key management. \\
\hline
\end{tabularx}

\end{table}

The topics \textit{scalable serving backends} and \textit{containerized GPU deployment} in Table \ref{tab:system_design_topics} show that there are repositories that emphasize API integration, resource management, performance optimization, Docker-based deployment, cloud services, and multi-GPU execution. This suggests that \textbf{LLM systems require architectural support for scalable and efficient serving in real deployment environments.} 
The topics \textit{Python/Conda pipeline organization}, \textit{dependency and script management}, and \textit{modular training pipelines} in the Table \ref{tab:system_design_topics} highlight the importance of organized environments, documentation, and evaluation integration. This indicates that \textbf{repositories using LLM serving frameworks require engineering structure to support maintainability.} 
A further finding of Table \ref{tab:system_design_topics} is that resource management is embedded in system design. 
The topics \textit{experiment tracking and GPU resource management} and \textit{CUDA and Tensor parallelism frameworks} show that repositories integrate experiment management, GPU utilization, CUDA kernels, parallelism, and efficient memory management into their architectures. 
This suggests that memory and resource management are important part of efficient LLM serving. This result further strengthen our findings of the usage of Memory management and parallelization methods presented on Table \ref{tab:coveredmethods}, and Figure \ref{fig:rq2.2_all}. Overall, the identified topics of Table \ref{tab:system_design_topics} suggest that repositories using LLM serving frameworks span multiple architectural layers, from low-level GPU execution to high-level orchestration and cloud deployment.



The findings of Table \ref{tab:system_design_topics} can benefit several stakeholder groups. \textbf{Developers of LLM serving frameworks can use them to improve support for containerized GPU deployment, scalable cloud infrastructure. 
MLOps and infrastructure engineers can use them to improve deployment pipelines, and resource management.} Finally, researchers can use these findings to better understand the broader architectural role of LLM serving frameworks in practical software systems and to identify important system-level challenges for future work.

\subsubsection{Which types of LLM-based software systems most commonly adopt each LLM serving frameworks, and how does this adoption vary across repository intent, technical focus, primary use case, and system design?}

This section presents the popularity of the topics among the repositories that use each of the studied serving frameworks. Table \ref{tab:summary_topics} presents the topics identified in repositories that use vLLM, SGLang, TensorRT-LLM, LMDeploy, and FlashInfer. 
As mentioned in Section \ref{rq3.1}, the summary categories are Repository Intent, Technical Focus, Primary Use Cases, and System Design. For each topic within each summary category, the table reports the topic name, topic number, and the number of repositories associated with each framework. Therefore, the table provides a comparative overview of how the studied serving frameworks are used in practice, what repository intents are associated with their use, what technical areas they are connected to, what application scenarios they support, and what system design patterns appear in repositories that adopt them.

\begin{table}[!htbp]
\centering
\caption{Summary categories and topic numbers across vLLM, SGLang, TensorRT-LLM, LMDeploy, and FlashInfer.}
\label{tab:summary_topics}
\scriptsize
\setlength{\tabcolsep}{4pt}
\renewcommand{\arraystretch}{1.2}

\resizebox{\textwidth}{!}{%
\begin{tabular}{|p{2.2cm}|c|c|c|c|c|c|c|}
\hline
\textbf{Summary categories} & \textbf{Topic Name} & \textbf{Topic No.} & \textbf{vLLM} & \textbf{SGLang} & \textbf{TensorRT-LLM} & \textbf{LMDeploy} & \textbf{FlashInfer} \\
\hline
\hline

\multirow{10}{*}{\centering Repository Intent}
& High-Performance GPU Serving & 0 & 129 & 53 & 34 & 9 & 36 \\
\cline{2-8}
& RL Frameworks & 1 & 60 & 44 & 2 & 1 & 2 \\
\cline{2-8}
& Mathematical Reasoning with RL & 2 & 31 & 16 & 2 & 1 & 1 \\
\cline{2-8}
& Retrieval and RAG Systems & 3 & 32 & 8 & 15 & 2 & - \\
\cline{2-8}
& Speech and Voice Applications & 4 & 24 & 6 & 4 & 1 & - \\
\cline{2-8}
& General ML and Generative AI Projects & 5 & 20 & 4 & 1 & 3 & - \\
\cline{2-8}
& Multimodal Vision Understanding & 6 & 13 & 11 & - & 1 & - \\
\cline{2-8}
& Chatbot and Agent Systems & 7 & 13 & 3 & - & 1 & - \\
\cline{2-8}
& Data Generation and Training Pipelines & 8 & 14 & 2 & - & - & - \\
\cline{2-8}
& Visual Grounding and Reasoning & 9 & 11 & 7 & - & - & - \\
\hline
\hline

\multirow{7}{*}{\centering Technical Focus}
& Model Training and Optimization & 0 & 301 & 121 & 15 & 21 & 14 \\
\cline{2-8}
& CUDA Parallelism and Kernel Optimization & 1 & 69 & 36 & 19 & 7 & 35 \\
\cline{2-8}
& Containerized API Deployment & 2 & 59 & 17 & 11 & 2 & 7 \\
\cline{2-8}
& Python Environment and Serving Workflows & 3 & 21 & 5 & 3 & 2 & 1 \\
\cline{2-8}
& Speech, OCR, and Multimedia Processing Pipelines & 4 & 17 & 5 & 11 & 1 & - \\
\cline{2-8}
& Hugging Face and RL Integration & 5 & 16 & 14 & - & - & - \\
\cline{2-8}
& Kubernetes-Based AI Deployment & 6 & 14 & 4 & - & - & - \\
\hline
\hline

\multirow{11}{*}{\centering Primary Use Cases}
& Multimodal Generation and Understanding & 0 & 69 & 31 & 17 & 11 & 3 \\
\cline{2-8}
& Efficient LLM Serving & 1 & 60 & 24 & 4 & 4 & 17 \\
\cline{2-8}
& Machine Learning Training and Usage & 2 & 51 & 17 & 5 & 6 & 5 \\
\cline{2-8}
& Cloud-Based AI Deployment & 3 & 44 & 10 & 12 & 1 & 3 \\
\cline{2-8}
& Long-Context QA and Benchmarking & 4 & 36 & 7 & 7 & 4 & 1 \\
\cline{2-8}
& Mathematical Reasoning Systems & 5 & 38 & 16 & 2 & 1 & 1 \\
\cline{2-8}
& Model Benchmarking and Fine-Tuning & 6 & 23 & 11 & 2 & 1 & 2 \\
\cline{2-8}
& RL Workflows & 7 & 24 & 18 & 2 & 1 & 1 \\
\cline{2-8}
& Agent-Based Systems & 8 & 20 & 5 & - & - & 4 \\
\cline{2-8}
& Retrieval and Recommendation Systems & 9 & 15 & 6 & - & - & - \\
\cline{2-8}
& GPU-Optimized Model Deployment & 10 & 13 & 7 & - & - & - \\
\hline
\hline

\multirow{9}{*}{\centering System Design}
& Scalable Serving Backends & 0 & 77 & 22 & 5 & 10 & 13 \\
\cline{2-8}
& Containerized GPU Deployment & 1 & 72 & 29 & 20 & 1 & 11 \\
\cline{2-8}
& Python/Conda Pipeline Organization & 2 & 47 & 17 & 5 & 3 & 2 \\
\cline{2-8}
& Dependency and Script Management & 3 & 35 & 13 & 1 & 1 & 10 \\
\cline{2-8}
& Microservices and Cloud Infrastructure & 4 & 26 & 4 & 4 & 3 & 1 \\
\cline{2-8}
& Experiment Tracking and GPU Resource Management & 5 & 21 & 20 & 4 & 1 & - \\
\cline{2-8}
& CUDA and Tensor Parallelism Frameworks & 6 & 9 & 6 & - & - & - \\
\cline{2-8}
& Modular Training Pipelines & 7 & 11 & 7 & - & - & - \\
\cline{2-8}
& Kubernetes-Based Cloud Infrastructure & 8 & 12 & 6 & - & - & - \\
\hline

\end{tabular}%
}

\end{table}

\textbf{Framework-level analysis}:
\textbf{vLLM appears in all identified topics in the Table \ref{tab:summary_topics} and has the highest count in all of topics compared to other frameworks.} 
Its most topic occurrences are in \textit{Model Training and Optimization}, \textit{High-Performance GPU Serving}, \textit{Scalable Serving Backends}, \textit{Containerized GPU Deployment}, \textit{CUDA Parallelism and Kernel Optimization}, \textit{Multimodal Generation and Understanding}, and \textit{Efficient LLM Serving}. One possible reason for this pattern is design of VLLM as a high-throughput and memory-efficient LLM inference and serving engine \cite{vllm_document}. 
Its documentation highlights memory management, optimized kernels, quantization, parallelism, OpenAI-compatible APIs, tool calling in AI Agents, reasoning parsers, and support for multimodal, embedding, retrieval, and reward models \cite{vllm_document}. 
Therefore, \textbf{the wide distribution of vLLM topics in Table \ref{tab:summary_topics} may reflect its role as a general-purpose serving infrastructure that is used both in low-level performance-oriented repositories and in higher-level application repositories.} 
SGLang also shows broad coverage across Table \ref{tab:summary_topics}, although its counts are generally lower than vLLM , which may be partly explained by the lower number of repositories adopting it. 
Its strongest topics include \textit{Model Training and Optimization}, \textit{High-Performance GPU Serving}, \textit{RL Frameworks}, \textit{CUDA Parallelism and Kernel Optimization}, \textit{Multimodal Generation and Understanding}, \textit{Containerized GPU Deployment}, \textit{Efficient LLM Serving}, and \textit{Scalable Serving Backends}. One possible reason for this results could be the focus of SGLang on low-latency and high-throughput serving for large language and multimodal models, using methods such as memory management, parallelism, quantization, and request scheduling \cite{sglang_document}. Besides, one possible reason for the presence of SGLang in reinforcement-learning, reasoning, agent topics could be the support by SGLang designed as a system for executing complex language-model programs involving multiple generation calls, control flow, agent control, logical reasoning, RAG pipelines\cite{zheng2024sglang}.

\textbf{TensorRT-LLM shows a more concentrated pattern than vLLM and SGLang across Table \ref{tab:summary_topics}.} Its highest occurrences appear in \textit{High-Performance GPU Serving}, \textit{Containerized GPU Deployment}, \textit{CUDA Parallelism and Kernel Optimization}, \textit{Multimodal Generation and Understanding}, \textit{Retrieval and RAG Systems}, \textit{Model Training and Optimization}, \textit{Cloud-Based AI Deployment}, and \textit{Speech, OCR, and Multimedia Processing Pipelines}. One possible reasons for this concentration is its support as an NVIDIA GPU-oriented inference framework for request scheduling, memory management, quantization \cite{TensorRT_LLM_document}, Docker containers, online serving, and multimodal applications. LMDeploy has lower counts than vLLM, SGLang, and TensorRT-LLM across Table \ref{tab:summary_topics}, but it appears across several application and deployment topics. Its strongest occurrences are in \textit{Model Training and Optimization}, \textit{Multimodal Generation and Understanding}, \textit{Scalable Serving Backends}, \textit{High-Performance GPU Serving}, \textit{CUDA Parallelism and Kernel Optimization}, \textit{Machine Learning Training and Usage}, \textit{Efficient LLM Serving}, and \textit{Long-Context QA and Benchmarking}. One possible reason for the occurrence of these topics is LMDeploy's support for compressing, deploying, and serving LLMs by providing a variety of memory management, parallelism, high-performance CUDA kernels, quantization, and request scheduling methods for resource-efficient LLM serving \cite{LMDeploy_document}. LMDeploy also documents support for many LLM and VLM families, including Llama, InternLM, Qwen, Baichuan, ChatGLM, and DeepSeek variants \cite{LMDeploy_document}, which support its appearance in multimodal and model-oriented topics. This may also explain why a broad range of LLMs are used with LMDeploy, as shown in the results of Section~\ref{rq2.3}. FlashInfer shows a specialized pattern in Table \ref{tab:summary_topics}. Its highest occurrences are in \textit{High-Performance GPU Serving}, \textit{CUDA Parallelism and Kernel Optimization}, \textit{Efficient LLM Serving}, \textit{Model Training and Optimization}, \textit{Scalable Serving Backends}, \textit{Containerized GPU Deployment}, and \textit{Dependency and Script Management}. One possible reason is of this concentrations is the FlashInfer’s design as a framework for kernel generator rather than an end-to-end application framework \cite{FlashInfer_document}. Efficient GPU attention kernels are essential for high-throughput and low-latency inference \cite{ye2025flashinfer}. Therefore, its stronger presence in GPU-serving and kernel-optimization topics, and weaker presence in application-level topics such as RAG, agents, speech, and recommendation, could be explained by its role as a low-level performance component.

\textbf{Summary category level and topic level analysis}:
\textbf{Within the Repository Intent category in Table~\ref{tab:summary_topics}, vLLM has the highest count across all topics, while SGLang is the second most frequently used framework in 90\% of the intent topics.} Both frameworks show their highest counts in the \textit{High-Performance GPU Serving}, \textit{RL Frameworks}, and \textit{Mathematical Reasoning with RL} repository intents. TensorRT-LLM is mainly associated with GPU serving and RAG-related repositories, whereas FlashInfer appears mostly in performance-oriented GPU serving topics, reflecting its role as a kernel-level serving framework rather than a full application framework. Also, LMDeploy has lower counts than vLLM and SGLang, but it appears across several intent topics, suggesting a smaller but relatively diverse observed usage. Besides, \textbf{Repository Intent \textit{High-Performance GPU Serving} is the most frequent topic across all studied frameworks.} This topic has the highest count for every framework which suggests that one important intent behind a considerable number of repositories is to improve or support efficient LLM inference on GPU-based infrastructures. Within the Technical Focus category in Table~\ref{tab:summary_topics}, \textit{Model Training and Optimization} is the dominant topic for vLLM, SGLang, TensorRT-LLM, and LMDeploy, while FlashInfer is most strongly associated with \textit{CUDA Parallelism and Kernel Optimization}. vLLM leads all topics, SGLang is broadly represented and close to vLLM in \textit{Hugging Face and RL Integration}, TensorRT-LLM is concentrated in \textit{CUDA parallelism and kernel-optimized} topic, and FlashInfer’s pattern reflects its GPU-kernel focus \cite{FlashInfer_document} by mostly being used in \textit{CUDA Parallelism and Kernel Optimization} topic. 

Within the Primary Use Cases category in Table~\ref{tab:summary_topics}, vLLM again has the highest count across all topics. TensorRT-LLM shows stronger occurrences in \textit{Multimodal Generation and Understanding} and \textit{Cloud-Based AI Deployment}, which could possibly be explained by its multimodal and deployment-oriented support \cite{TensorRT_LLM_document}. In the System Design category in Table~\ref{tab:summary_topics}, vLLM has the highest occurrence in all topics, especially in \textit{Scalable Serving Backends} and \textit{Containerized GPU Deployment}, this may indicate its association with scalable and GPU-based deployment architectures. TensorRT-LLM is mainly concentrated in \textit{Containerized GPU Deployment}. Also, LMDeploy appears with highest counts across \textit{Scalable Serving Backends}. FlashInfer is mostly associated with \textit{Scalable Serving Backends}, \textit{Containerized GPU Deployment}, and \textit{Dependency and Script Management}. Overall, \textbf{the results show that scalable backend design and containerized GPU deployment are the dominant system-design patterns, while the more specialized topics, including \textit{CUDA and Tensor Parallelism Frameworks}, \textit{Modular Training Pipelines}, and \textit{Kubernetes-Based Cloud Infrastructure}, appear only for vLLM and SGLang in the studied repositories.}

Table \ref{tab:summary_topics} can benefit several stakeholder groups because it connects each studied serving framework with the types of repositories in which it appears. Developers and maintainers of LLM serving frameworks can use the results of Table \ref{tab:summary_topics} to better understand the practical contexts in which their frameworks are adopted. For example, the table can help them identify whether their framework is mostly used for high-performance GPU serving, reinforcement-learning workflows, multimodal applications, cloud deployment, or scalable backend systems. \textbf{This information can guide framework developers in improving APIs, documentation, examples, and integration support for the repository types in which their frameworks are most frequently used. It can also help them identify application or system-design contexts where their framework has limited observed adoption and may require better support.} Application developers who use LLM serving frameworks can also benefit from these results. Table \ref{tab:summary_topics} provides empirical evidence about which frameworks are commonly associated with specific application scenarios, such as multimodal generation and understanding, long-context question answering, reinforcement-learning workflows, and so on. Therefore, \textbf{developers can use the Table \ref{tab:summary_topics} as a practical reference when selecting a serving framework for their target application. This can help reduce the effort required to compare multiple frameworks and can support more informed framework selection.}

MLOps, and infrastructure engineers can use the results of Table \ref{tab:summary_topics} to identify the system-level requirements that commonly appear in repositories using LLM serving frameworks. \textbf{These commonly used system-level topics suggest that LLM serving frameworks are often embedded in broader deployment and infrastructure pipelines rather than used only as isolated inference tools.} As a result, \textbf{infrastructure engineers can use these findings to improve deployment pipelines, resource management, containerization support, and cloud orchestration for LLM-based systems. Researchers in software engineering and AI systems can use the findings of Table \ref{tab:summary_topics} to identify important research directions such as system design challenges that the developers may face when using LLM serving frameworks.} Finally, technology decision-makers adopting LLM systems can use Table \ref{tab:summary_topics} to better understand the practical landscape of LLM serving frameworks. The table can support decisions about which frameworks to evaluate, which infrastructure capabilities to prioritize, and which engineering skills may be needed for deployment. In this way, the results can help organizations make more evidence-based decisions about LLM serving infrastructure.




\begin{tcolorbox}[breakable]
\textbf{Finding 3.} Repositories using LLM serving frameworks cover diverse software-system types, including high-performance serving, RL, RAG, and multimodal systems. Their designs span multiple architectural layers, from low-level GPU execution and resource management to scalable backends, containerized deployment, and cloud orchestration. Topic adoption is uneven: vLLM has the broadest adoption across all four dimensions, while SGLang is also broadly adopted across several contexts. TensorRT-LLM is concentrated in GPU-oriented and deployment-heavy contexts. LMDeploy shows smaller but diverse adoption across training, efficient-serving, long-context, and scalable-backend topics. FlashInfer is mainly specialized around GPU-serving, kernel optimization, and serving-backend infrastructure.
\end{tcolorbox}

\section{Related Works}
\label{relatedworks}
This section presents and discusses the related literature.

\subsection{Efficient LLM Inference and Serving}
Efficient inference and serving of LLMs are important for understanding how LLM-based systems can be deployed with practical latency, throughput, memory use, and deployment efficiency \cite{miao2025towards,pan2025survey}. Prior studies have examined efficient and resource-efficient LLMs from broad technical perspectives. Bai et al.\cite{1bai2024beyond} review resource-efficient LLM techniques across architecture design, pre-training, fine-tuning, inference, and system design, while Wan et al. \cite{9wan2023efficient} organize efficient LLM research into model-centric, data-centric, and framework-centric directions. In a broader foundation-model context, Xu et al. \cite{13xu2024survey} study resource-efficient architectures, algorithms, and systems, whereas Zhou et al. \cite{14zhou2024survey} focus on efficient LLM inference through data-level, model-level, and system-level optimization. Furthermore, Miao et al. \cite{miao2025towards} review efficient LLM serving methods from both algorithmic and system-level perspectives and compare representative open-source GPU-based LLM serving systems. Pan et al. \cite{pan2025survey} review LLM inference systems through the inference stack, including request processing, model execution, batching, scheduling, kernel design, and memory management. These studies provide useful background for understanding efficient LLM methods, inference, frameworks, system design, and the technical space of LLM serving. However, their main focus is to review, classify, organize, and conceptually compare existing techniques and systems, rather than empirically examine how LLM serving frameworks, serving methods, framework combinations, and system designs are used in open-source GitHub repositories. Our study complements this line of work by empirically analyzing these aspects in practice.

\subsection{Algorithmic Techniques for Efficient LLM Inference}
Algorithmic techniques are important part of efficient LLM inference because they aim to reduce memory use and computation cost during model execution \cite{10wang2024model,15zhu2024survey}. Prior work has examined algorithmic and model-level methods for efficient LLM inference. Wang et al. \cite{10wang2024model} study LLM compression and inference acceleration from an algorithmic perspective that reduce memory and computation costs during inference. Also, Zhu et al. \cite{15zhu2024survey} review LLM model compression techniques, including quantization, network pruning, together with related metrics, benchmarks, and open challenges. Furthermore, Xia et al. \cite{12xia2024unlocking} study speculative decoding, which aims to accelerate autoregressive LLM inference, by organizing existing methods into a taxonomy, and comparing representative approaches for inference acceleration. The paper also highlights open challenges such as balancing accuracy and latency, improving batched speculative decoding, and integrating speculative decoding with other efficient inference frameworks and techniques such as vLLM and continuous batching. Both surveys show that efficient inference is closely connected to reducing model size, memory demand, and computational cost while preserving the general capabilities of LLMs, and speculative decoding is one of the efficient inference methods that may be implemented or exposed by LLM serving frameworks. Although \cite{10wang2024model,15zhu2024survey} provide useful background for our study because several of these techniques can appear as serving methods supported by LLM serving frameworks, based on the provided summaries, these works mainly organize algorithmic and compression methods, rather than empirically examining how such methods are adopted, combined, or used in open-source GitHub repositories. Similarly, Xia et al. \cite{12xia2024unlocking} focus on the method itself, while our study complements these works by analyzing how LLM serving methods, including speculative decoding, framework combinations, and system designs are used in open-source GitHub repositories.

\subsection{System-Level Optimization for Efficient LLM Serving}
System-level optimization is important for practical LLM serving because inference cost and performance can be affected by runtime routing, memory management, hardware allocation, and framework-level execution strategies \cite{3ding2024hybrid,4gao2024cost,5jiang2025demystifying,11xia2023flash}. \cite{3ding2024hybrid,4gao2024cost,5jiang2025demystifying,11xia2023flash} illustrated how such system-level decisions can improve serving efficiency from different perspectives, including query routing, KV-cache reuse, heterogeneous GPU deployment, and sparse GPU-based inference \cite{3ding2024hybrid,4gao2024cost,5jiang2025demystifying,11xia2023flash}. System-level methods and approaches have also been proposed to improve the cost efficiency of LLM serving at runtime. One approach is query routing, where Ding et al. \cite{3ding2024hybrid} study how to reduce inference cost by sending easier queries to a smaller model and harder queries to a larger model while keeping response quality close to that of the larger model. Another approach is memory-aware serving for multi-turn conversations, where Gao et al. \cite{4gao2024cost} propose CachedAttention to reuse historical KV caches instead of recomputing them for each new turn. Other system-level studies focus on hardware-aware and framework-level efficiency for LLM serving. Jiang et al. \cite{5jiang2025demystifying} examine cost-efficient serving over heterogeneous cloud GPUs and jointly optimizes GPU composition, deployment configuration, and workload assignment under budget and availability constraints. Xia et al. \cite{11xia2023flash} introduce Flash-LLM, a GPU-based library that improves large generative model inference by supporting sparse matrix multiplication on tensor cores. These works address different runtime inefficiencies and show that serving efficiency can depend on reducing unnecessary computation, deployment decisions, GPU resources, memory access, and framework-level execution strategies. They are related to our study because they represent concrete serving methods and highlight system-level concerns that may appear in LLM serving frameworks or open-source systems. However, they study specific optimization methods rather than how such methods are adopted, combined, and used across GitHub repositories, and they do not empirically analyze open-source GitHub adoption patterns, framework combinations, or repository-level system designs, which are the main focus of our study.

\subsection{LLM Serving Frameworks and Deployment Infrastructure}
Serving frameworks and deployment infrastructure are important for moving model-serving systems from experimental settings to practical production environments \cite{2beck2025evaluation,8shan2025aibrix}. \cite{2beck2025evaluation} evaluates open-source model serving tools through feature comparison, runtime experiments, and experience from real-world ML projects. The study shows that serving tools differ in performance, setup difficulty, framework compatibility, scalability, and production requirements, and that no single tool is best for all serving scenarios. This work also treats model serving as an important part of production ML systems and provides practical guidance for selecting serving environments. More recent work has moved toward production-grade LLM-serving infrastructure. Shan et al. \cite{8shan2025aibrix} introduce AIBrix as a cloud-oriented LLM inference infrastructure designed to improve large-scale deployment, performance, scalability, and cost efficiency. The framework integrates several serving-related components, including LoRA management, autoscaling, routing, unified runtime support, distributed KV cache, hybrid orchestration, heterogeneous GPU optimization, and diagnostic tools. This work presents a LLM-serving framework and discusses several serving methods used in production-oriented settings. However, Beck et al. \cite{2beck2025evaluation} focuse mainly on general-purpose ML model-serving tools and do not examine how LLM serving frameworks and serving methods are adopted or combined across open-source GitHub repositories, while Shan et al. \cite{8shan2025aibrix} study one framework rather than empirically analyzing how LLM serving frameworks, serving methods, framework combinations, and system designs are used across open-source GitHub repositories.

\section{Discussion and Implications}
\label{discussion}
Our findings show that efficient  serving in open-source repositories is shaped by the interaction between framework capabilities, serving-method availability,  types, and repository-level system goals. Developers frequently rely on methods that address common serving bottlenecks, particularly memory management and parallel execution, while framework-specific patterns show more specialized uses, such as the stronger role of kernel- and attention-level optimization in FlashInfer repositories. The results also show that multi-framework usage remains limited, suggesting that most repositories rely on a single studied serving framework. However, when frameworks are combined, recurring combinations such as FlashInfer + vLLM, SGLang + vLLM, and FlashInfer + SGLang connect complementary capabilities across the serving stack, including orchestration, scheduling, memory management, parallel execution, and attention- or kernel-level optimization.


Our results provide more specific implications for framework maintainers, developers, and researchers. The RQ1 results show that Additional Parallel Computation, Memory Management, and Network Pruning are the most frequently observed serving-method categories, while FlashInfer repositories show a stronger concentration around Kernel Fusion. This suggests that maintainers can improve documentation and examples around frequently used optimization paths, such as memory-aware serving, parallel execution, and kernel-level optimization. The RQ2 results further show that some repositories combine multiple methods within the same framework and that recurring framework combinations, including FlashInfer + vLLM, SGLang + vLLM, and FlashInfer + SGLang, connect complementary capabilities across scheduling, memory management, parallel execution, and attention- or kernel-level optimization. Therefore, maintainers could provide integration examples, compatibility notes, and configuration templates for common multi-method and cross-framework settings rather than documenting each method only in isolation. The RQ3 results also show that LLM serving frameworks are used across diverse repository-level contexts, such as mathematical reasoning with RL, multimodal generation, multimedia processing pipelines, and microservice-based infrastructure, suggesting that developers and researchers can use these results to select and study serving frameworks with respect to both model characteristics and system-level deployment goals.


\section{Threats to Validity}
\label{threats}
This section discusses the main threats to the validity of our study and the strategies used to reduce their possible effects. We organize the threats into construct, internal, and external validity.

\subsection{Construct Validity}
One possible construct validity threat is related to the classification of LLM serving methods. We adopted an established taxonomy of LLM serving methods and studied the official documentation of each selected framework to identify the APIs associated with each method. However, some framework-specific methods may not exactly match the existing taxonomy categories. To mitigate this threat, we carefully reviewed the framework documentation and, when a method did not directly fit the taxonomy, we added additional sub-subcategories instead of forcing it into an inaccurate class.

A further construct validity threat is related to the representation of repository characteristics for clustering and topic modeling. Repository descriptions may be incomplete, and generated summaries may not fully capture the actual intent, technical focus, use case, or system design of each repository. To address this threat, we generated repository summaries using three complementary metadata sources: README files, about sections, and repository keywords.

\subsection{Internal Validity}
One possible threat concerns possible errors during the aggregation of scripts into repository-level results. A single repository may contain multiple Python scripts, and each script may include several framework or method APIs. If raw script occurrences were counted directly, the adoption of some frameworks or methods could be overestimated. To reduce this threat, we counted unique repositories rather than raw script occurrences when reporting framework adoption, method usage, and framework combinations. This helped ensure that repositories with repeated API occurrences did not dominate the results.

\subsection{External Validity}
One possible threat is related to the focus on Python scripts. Prior studies identify Python as the most used programming language for developing LLM-based software systems \cite{twist2025study}. Thus, we believe that the results of this study are applicable to a considerable portion of LLM-serving systems. 

Another threat is the focus of this study on open-source GitHub repositories. We reduced this threat by clearly defining the scope of the study as open-source repositories and by following prior empirical software-engineering studies that use GitHub as a source of software systems \cite{kalliamvakou2014promises}.

\section{Conclusion and Future Work}
\label{conclusion}

As LLM are increasingly integrated into practical software systems, understanding how they are served efficiently in practice has become an important software engineering concern. However, empirical evidence on how serving frameworks and efficient serving methods are adopted and combined in open-source repositories remains limited. To address this gap, this study analyzed five  serving frameworks, their supported efficient serving methods, method and framework combinations, the serving frameworks adopted for different types of LLM, and the repository-level characteristics of software systems that adopt these frameworks in terms of repository intent, technical focus, primary use case, and system design. The results show that vLLM is the most popular and adopted framework, while parallel computation, memory management, and network pruning are among the most frequently observed efficient serving-method categories. The results also show that developers combine multiple serving methods within frameworks, whereas multi-framework usage remains limited. In multi-framework repositories, vLLM has the largest absolute co-usage count, while FlashInfer has the highest co-usage ratio, indicating that it is more often adopted alongside other serving frameworks. These recurring combinations connect complementary capabilities across the serving stack, including scheduling, memory management, parallel execution, and attention- or kernel-level optimization. Furthermore, the findings show that serving-framework adoption differs across  model families, modalities, sizes, specializations, and deployment contexts, and that serving frameworks are used across diverse repository-level contexts, including mathematical reasoning with RL, multimedia processing pipelines, multimodal generation, and microservice-based infrastructure. Overall, these findings provide empirical evidence for researchers, developers, and framework maintainers about how  serving frameworks and efficient serving methods are used in real open-source software systems. Future work can extend this study by examining how AI-agent systems that rely on LLM can be served efficiently, and by conducting empirical studies to identify the practical challenges developers face when efficiently deploying and serving LLM in real-world software systems.

\section{Declaration}
\textbf{Ethical approval}: Not applicable. This study does not involve any human participants or animals. Therefore, we do not require ethical approval.

\textbf{Informed consent}: Not applicable. This study does not involve a survey or human participants. Therefore, we do not require informed consent.

\textbf{Author contributions}: Forough Majidi and Mehdi Morovati proposed the study idea and approach, revised the methodology, interpreted the results, developed the Python scripts, and contributed to drafting and revising the manuscript, with Forough Majidi taking the lead across these activities. Foutse Khomh and Heng Li contributed to refining the study idea, approach, and methods. They also improved the manuscript by restructuring the paper, identifying and addressing writing and presentation issues, and proofreading the manuscript.

\textbf{Data availability}: All data used for analysis, as well as the scripts and the analysis results, are provided in \cite{replicationPackage}.

\textbf{Conflict of interest}: The authors declared that they have no conflict of interest.

\textbf{Clinical Trial Number}: Not applicable

%
%
\bibliographystyle{IEEEtran}
\bibliography{bibliography}


\end{document}